\documentclass[preprint]{aastex}
\usepackage{amsmath}
\newcommand\OVI{O~{\scriptsize VI}}
\newcommand\OVII{O~{\scriptsize VII}}
\newcommand\OVIII{O~{\scriptsize VIII}}
\newcommand\CIV{C~{\scriptsize VI}}
\newcommand{\ov}{O~\textsc{v}}
\newcommand{\ovi}{O~\textsc{vi}}
\newcommand{\ovii}{O~\textsc{vii}}
\newcommand{\oviii}{O~\textsc{viii}}

\newcommand\civ{C~\textsc{iv}}
\newcommand\ciii{C~\textsc{iii}}

\newcommand\caii{Ca~\textsc{ii}}
\newcommand\hi{H~\textsc{i}}

\newcommand{\Ioviii}{\ensuremath{I_\mathrm{O\,VIII}}}
\newcommand{\Iovii}{\ensuremath{I_\mathrm{O\,VII}}}
\newcommand{\Iovi}{\ensuremath{I_\mathrm{O\,VI}}}
\newcommand{\Iciv}{\ensuremath{I_\mathrm{C\,IV}}}
\newcommand{\rovi}{\ensuremath{r_\mathrm{O\,VI}}}
\newcommand{\rovii}{\ensuremath{r_\mathrm{O\,VII}}}
\newcommand{\roviii}{\ensuremath{r_\mathrm{O\,VIII}}}

\newcommand{\nhi}{\ensuremath{N_\mathrm{H\,I}}}
\newcommand{\nh}{\ensuremath{N_\mathrm{H}}}

\newcommand{\angstrom}{~\ensuremath{\mbox{\AA}}}
\newcommand{\flux}{~\textnormal{photons}~\textnormal{cm}^{-2}~\textnormal{s}^{-1}
~\textnormal{sr}^{-1}}
\newcommand{\lu}{~\textnormal{LU}}
\newcommand{\eflux}{~\textnormal{photons}~\textnormal{cm}^{-2}~\textnormal{s}^{-1}
~\textnormal{sr}^{-1}~\textnormal{keV}^{-1}}
\newcommand{\ergflux}{~\textnormal{ergs}~\textnormal{cm}^{-2}~\textnormal{s}^{-1}}
\newcommand{\erglu}{~\textnormal{ergs}~\textnormal{s}^{-1}}
\newcommand{\countrate}{~\textnormal{counts}~\textnormal{s}^{-1}~\textnormal{arcmin}^{-2}}
\newcommand{\uip}{~\textnormal{cm}^{-3}~\textnormal{s}}
\newcommand{\uden}{~\textnormal{cm}^{-3}}

\newcommand{\pcms}{~\textnormal{cm}^{-2}}
\newcommand{\utau}{~\textnormal{s}~\textnormal{cm}^{-3}}
\newcommand{\uem}{~\textnormal{cm}^{-6}~\textnormal{pc}}
\newcommand{\K}{~\textnormal{K}}
\newcommand{\kev}{~\textnormal{keV}}
\newcommand{\pc}{~\textnormal{pc}}
\newcommand{\kpc}{~\textnormal{kpc}}
\newcommand{\ergs}{~\textnormal{ergs}}
\newcommand{\ergpers}{~\textnormal{erg}~\textnormal{s}^{-1}}
\newcommand{\xflux}{~\textnormal{erg}~\textnormal{s}^{-1}~\textnormal{kpc}^{-2}}
\newcommand{\yr}{~\textnormal{yr}}
\newcommand{\acrate}{~\textnormal{M}_\odot~\textnormal{yr}^{-1}}
\newcommand{\ifrate}{~\textnormal{M}_\odot~\textnormal{yr}^{-1}~\textnormal{kpc}^{-2}}
\newcommand{\um}{~\mu\textnormal{m}}

\newcommand\xmm{\textit{XMM-Newton}}
\newcommand{\Suzaku}{\textit{Suzaku}}
\newcommand{\ROSAT}{\textit{ROSAT}}
\newcommand{\IRAS}{\textit{IRAS}}
\newcommand{\FUSE}{\textit{FUSE}}
\newcommand{\SPEAR}{\textit{SPEAR}}
\newcommand{\Chandra}{\textit{Chandra}}

\begin{document}
\title{Determining The Galactic Halo's Emission Measure from UV and X-ray Observations}
\author{Shijun Lei, Robin L. Shelton and David B. Henley}
\affil{Department of Physics and Astronomy, University of Georgia, Athens, GA 30602}
\email{sjlei@physast.uga.edu}

\begin{abstract}
We analyze a pair of \Suzaku\ shadowing observations in order to determine the X-ray spectrum of the
Galaxy's gaseous halo. Our data consist of an observation toward an absorbing filament in the southern
Galactic hemisphere and an observation toward an unobscured region adjacent to the filament. We
simultaneously fit the spectra with models having halo, local, and extragalactic components. The
intrinsic intensities of the halo \ovii\ triplet and \oviii\ Lyman $\alpha$ emission lines are
$9.98^{+1.10}_{-1.99}\lu$ (line unit;$\flux$) and $2.66^{+0.37}_{-0.30}\lu$, respectively.  These results
imply the existence of hot gas with a temperature of $\sim$$10^{6.0}\K$ to $\sim$$10^{7.0}\K$ in the
Galactic halo. Meanwhile, \FUSE\ \ovi\ observations for the same directions and \SPEAR\ \civ\
observations for a nearby direction indicate the existence of hot halo gas at temperatures of
$\sim$$10^{5.0}\K$ to $\sim$$10^{6.0}\K$. This collection of data implies that the hot gas in the Galactic
halo is not isothermal, but its temperature spans a relatively wide range from $\sim$$10^{5.0}\K$ to
$\sim$$10^{7.0}\K$. We therefore construct a differential emission measure (DEM) model for the halo's
hot gas, consisting of two components. In each, $d\textnormal{EM}/d\log T$ is assumed to follow a
power-law function of the temperature and the gas is assumed to be in collisional ionizational
equilibrium. The low-temperature component (LTC) of the broken power-law DEM model covers the
temperature range of $10^{4.80}-10^{6.02}\K$ with a slope of 0.30 and the high-temperature component
(HTC) covers the temperature range of $10^{6.02}-10^{7.02}\K$ with a slope of $-2.21$. We compare our
observations with predictions from models for hot gas in the halo. The observed spatial distribution of
gas with temperatures in the range of our HTC is smoother than that of the LTC.
We thus suggest that two types of sources contribute to our broken power-law
model. We find that a simple model in which hot gas accretes onto the Galactic halo and cools radiatively
cannot explain both the observed UV and X-ray portions of our broken power-law model. It
can, however, explain the intensity in the \Suzaku\ bandpass if the mass infall rate is
$1.35\times10^{-3}\ifrate$. The UV and X-ray intensities  and our broken power-law model can be well
explained by hot gas produced by supernova explosions or by supernova remnants supplemented by a
smooth source of X-rays.
\end{abstract}


\section{Introduction}
Not only does hot gas ($T>10^5\K$) reside in our galaxy's disk, but it resides in the halo. (Here we
use the X-ray astronomy convention which defines the halo as the region above the majority of the
Galaxy's \hi, thus above a height of $z\sim150-200\pc$ given the parameterization of the \hi\ distribution
by \cite{ferriere98b} and \cite{dickey90}, although other conventions would call the lower part of this
region the thick disk.) Utraviolet and X-ray observations indicate that the high-latitude sky is covered by
hot gas. Absorption by Galactic \ovi\ ions, tracers of $T\sim3\times10^5\K$ gas, is seen in all of the
$Far~Ultraviolet~Spectroscopic~Explorer$ (\FUSE) halo survey spectra of sight lines that transit the
halo and have high signal-to-noise data \citep{wakker03}. $1/4\kev$ X-rays, tracers of  $T\sim10^6\K$
gas, are also seen in all directions, but some of these X-rays are produced locally, either in the Local
Bubble (LB) or in the heliosphere, and by external galaxies. After the local and extragalactic contribution
are subtracted, X-rays are found to come from most, if not all, high latitude directions \citep{snowden98}.
Such a large covering fraction does not require that the hot gas forms a smooth layer. In fact, maps of
\ovi\ column density and $1/4\kev$ brightness show a mottled or lumpy distribution. In comparison,
maps of $3/4\kev$ brightness are far smoother, with the exception of the North Polar Spur/Loop I region
which is bright in both 1/4 and $3/4\kev$ X-rays \citep[see maps in ][]{snowden97}. 

The height of the hot gas has been found from the \ovi\ column density data. The average density of
\ovi\ ions falls off exponentially with height above the plane and has scale-heights of 4.6 and
$3.2\kpc$ for northern and southern Galactic hemispheres, respectively \citep{bowen08}. Although
it is not possible to calculate the hot gas scale-height from observations of diffuse X-ray emission, it is
possible to determine whether or not X-rays are produced beyond clouds of neutral or molecular
interstellar gas. Such analyses, dubbed ``shadowing" analysis, find $1/4\kev$ X-rays originating beyond
clouds at heights of
$\sim$$160\pc$ (southern filament: \citealp{wang95}, \citealp{shelton07}, distance from
\citealp{penprase98}),
$\sim$$200\pc$ (Draco cloud: \citealp{burrows91}, \citealp{snowden91}, distance from
\citealp{lilienthal91}),
$\sim$$285\pc$ (Ursa major cloud: \citealp{snowden94b}, distance from
\citealp{benjamin96}), and 
$\sim$$1.5\kpc$ (Complex M clouds: \citealp{herbstmeier95}, distance from
\citealp{danly93}). 
One analysis of the shadowing filament in the Southern hemisphere
($z=160\pm20\pc$) reported both the intensity of ultraviolet photons emitted by \ovi\ ions and the
intensity of $1/4\kev$ X-rays emitted by hotter gas. Coupling the UV and soft X-ray regimes was fruitful;
it led to the realization that the radiation by hot gas in the halo accounts for a significant fraction of the
energy injected into the Galaxy at the Sun's galactocentric radius \cite[$\sim6\times10^{38}\xflux$ vs. an
energy injection rate of $\sim8\times10^{38}\xflux$ due to supernova (SN) and pre-SN winds;][]{shelton07}.

Higher energy ($0.3-2\kev$) \Suzaku\ shadowing observations were made for the same southern
hemisphere filament as was observed with \FUSE\ and the \ROSAT\ All Sky Survey (RASS).
\citeauthor{henley08} (2008, hereafter Paper I) processed the raw \Suzaku\ data and extracted
both the local and halo spectra, but primarily used the data in order to analyze the LB and
compare with solar wind charge exchange (SWCX) contaminated \xmm\ data for the same
directions. Here, we combine the \Suzaku\ spectrum of the halo with \ROSAT\ $1/4\kev$ and \FUSE\
\ovi\ shadowing data in order to create the first long-baseline spectrum of a single region of the Galactic
halo. We extend our spectrum to $1550\angstrom$ using \civ\ data for nearby, but not coincident,
pointings taken by instruments on the
$Spectroscopy~of~Plasma~Evolution~from~Astrophysical~Radiation$ (\SPEAR) satellite. We
compare the X-ray portion of the long-baseline spectrum with model spectra for
collisional ionizational equilibrium (CIE) and non-equilibrium ionization (NEI) plasmas. We test
isothermal models as well as
models that have two thermal components. Although the X-ray data are well fitted by models having two
thermal components, the combined UV and X-ray spectrum is not. Because the long baseline spectrum
requires some EM at lower temperatures, we fit it with a differential emission measure
(DEM) function. Such a broad spectrum is a powerful tool for testing phenomenological models for hot
gas in the Galactic halo. We compare our results with three models, two of which are phenomenological
(accreting gas and supernova remnant(SNR)), while the third assumes that the quantity and temperature
of hot gas varies smoothly with height above the plane \citep{yao07a}. We find that the intensity and
spectrum of an SNR are consistent with the observations, that the predictions for a simple
model of accreted gas under-produce the UV intensities relative to the X-ray intensities, and that the
geometrical model must be modified in order to account for the \ovi\ intensity seen in our direction.

The observations are described in Section 2 The basic assumptions of the analysis method are described
in Section 3. Section 4 discusses our tests of the most basic models, those using isothermal and two-temperature
thermal spectra for plasmas in or approaching CIE. In Section 5, we measure the halo's intrinsic \ovii\ and
\oviii\ line intensities and use them, together with the intrinsic \civ\ and \ovi\ intensities obtained from
\SPEAR\ and \FUSE\ observations, to estimate the halo's EM distribution as a
function of temperature, $T$, in the range $T\sim10^{5.0}-10^{7.0}\K$. We follow up this preliminary
investigation by testing various possible DEM models in Section 6. Our best-fitting broken-power-law model
is given in Section 6.2. The limitations and the physical implications of our modeling are discussed in Section 7
followed by a summary in Section 8.

\section{Observations and Data Reduction}
Shadowing studies require a set of  observations, with one observation toward a molecular
cloud or filament and one nearby. In this study, the on-filament observation was toward a dense knot
in the filament described by \cite{penprase98} ($l=278.65\degr$, $b=-45.30\degr$). The off-filament
observation was toward an unobscured direction approximately 2\degr\ away ($l=278.71\degr$,
$b=-47.07\degr$). The on- and off-filament \Suzaku\ spectra analyzed here are identical to those
described in Paper I. In what follows, we just use the data from \Suzaku's back-illuminated XIS1 chip,
which is more sensitive at lower energies than the three front-illuminated chips. Details of the \Suzaku\
observations and data reduction were given in Paper I. Specifically, point sources with $0.2-4.5\kev$
fluxes above $5\times10^{-14}\ergflux$ were excluded using the data from the prerelease of the second
\xmm\ Serendipitous Source Catalogue. 

The \FUSE\ \ovi\ intensities \citep{shelton03, shelton07} and the \ROSAT\ R12 and R45 count rates
\citep{snowden97} for the same directions are taken from the existing literature, and the readers are
referred to these papers for information on the observations and data reduction. The \SPEAR\ \civ\
intensities for the sight lines near ours were given to us by J. Kregenow (2006, private communication).

\section{Basic Assumptions and Analysis Method}
The total diffuse X-ray emission along most high latitude lines of sight is generally attributed to three
basic sources, namely  the LB, the extragalactic power-law (EPL) background due to unresolved distant
active galactic nuclei (AGNs), and the Galactic halo \citep{snowden98, kuntz00}. Each source is then
modeled with one or 
more components in our multi-component model fitting to the observational data. Throughout this
paper we adopt a thermal plasma component in collisional ionization equilibrium (CIE) for the LB
emission. We use a power law with a photon index of $1.46$ to model the extragalactic background
\citep{chen97}. The normalization of the power-law model is a free parameter to be determined by the
fitting, and we obtain $\sim$$11$ and $\sim$$8\eflux$ at $1\kev$ for the on- and off-filament lines of
sight respectively. In this paper, we focus on the hot gas in the Galactic halo, and so we investigate
various models for the halo component. Since the same models for the LB and EPL components are
always included in our fitting, for simplicity, we hereafter name the entire multi-component model
only after the name of the model(s) for the halo component(s). In all of our models the LB component is
unabsorbed, and the halo and EPL components are subject to absorption. Besides the three basic X-ray
emitting sources modeled in our spectral analysis, two other sources, i.e., SWCX and the X-ray emitting
stellar population, are known to produce possible contamination in our \Suzaku\ observations. But as
we shall argue in Section 7.1, the contamination from these two sources is small and can be safely ignored.

As the \Suzaku\  XIS1 is not well calibrated below $0.3\kev$, we truncate our \Suzaku\ spectra at
$0.3\kev$. The final \Suzaku\ spectra we analyze cover the energy range of $0.3$-$5.5\kev$. We also
omit the data between $1.3$-$2.3\kev$, where the \Suzaku\ spectra are contaminated by instrumental
emission lines. \ROSAT\ spectra are also available for the same sight lines from the RASS. Although the
\ROSAT\ spectra are of much lower spectral resolution than the \Suzaku\ spectra, they still make a good
supplement to the \Suzaku\ spectra because they extend the energy range down to $\sim$$~0.1\kev$.
We therefore include in our analysis the \ROSAT\ R12 data, which cover the $\sim$$0.1-0.284\kev$
energy band. We fit to the \Suzaku\ and \ROSAT\ R12 (hereafter \Suzaku+\ROSAT) spectra jointly. The
higher energy \ROSAT\ bands are not included because these energy bands overlap with the \Suzaku\
spectra, which are of much higher signal-to-noise and energy resolution. We use XSPEC to generate
most of our spectral models and fit them to the observed spectra. Following the argument in Paper I, we
use the Astrophysical Plasma Emission Code (APEC) v1.3.1 (Smith et al. 2001) to simulate thermal
models that we fit to the \Suzaku\ spectra and the Raymond \& Smith (RS) code \citep{raymond77,
raymond91} to
make thermal models that we fit to the \ROSAT\ R12 data. During the fitting, the parameters of the RS
model components for the \ROSAT\ data are tied to the parameters of the corresponding APEC model
components for the \Suzaku\ spectra. In all cases we use the $\mathtt{phabs}$ model for the absorption
(where typewriter font denotes XSPEC commands). The interstellar medium (ISM) abundance table
from \cite{wilms00} is  used for the chemical abundances of both the thermal plasma and the absorbing
media \citep[see][]{henley07}.

Following Paper I, we take $\nh=9.6\times10^{20}$ and $1.9\times10^{20}\pcms$ \citep{henley07} as our
``standard" values for the on- and off-filament hydrogen column densities throughout the paper. Those
values were
estimated using the $100\um$ intensities, $I_{100}$, from the all-sky \IRAS\ maps of \cite{schlegel98}
and the $I_{100}$-to-$\nh$ conversion relation for the southern Galactic hemisphere given in
\cite{snowden00}. The on-filament column density is consistent with the value derived from the color
excess of the filament, $E(B-V)= 0.17\pm 0.05$ \citep{penprase98}, which yields
$\nh=(9.9\pm2.9)\times10^{20}\pcms$ when scaled using the conversion relation given by
\cite{bohlin78}. Measurements of $\nhi$, made from observations of its 21~cm intensity, are provided
by the Leiden-Argentine-Bonn (LAB) Galactic \hi\ Survey \citep{kalberla05}, which gives
$\nhi=(5.1\pm0.1)\times10^{20}$ and $(2.0\pm0.1)\times10^{20}\pcms$ for the on- and off-filament
directions respectively. This off-filament $\nhi$ value is similar to the \nh\ value derived from the
$I_{100}$ intensity, but the difference between the on-filament \nh\ and \nhi\ measurements is
significant. This difference is probably due to molecular hydrogen. 

For the uncertainty in the on-filament column density, we use the value derived from the color excess
measurement of \cite{penprase98}, yielding $\nh=(9.9\pm2.9)\times10^{20}\pcms$. In order to account
for the uncertainty in the off-filament column density, we take $\nh=2.0\times10^{20}\pcms$ as the upper
limit and $\nh=0.5\times10^{20}\pcms$ as the lower limit, following \cite{shelton07}. The two
models upon which our conclusions are based are tested
for various on- and off-filament column densities. In the upcoming spectral fits, we find that the
uncertainties in the on- and off-filament X-ray absorbing column densities have a limited effect on the
general results. More details of the influence of these uncertainties on the specific models are given
in subsections 4.1 and 6.2.

\section{Isothermal and Two-Temperature Halo Models}
\subsection{Collisional Ionizational Equilibrium Models}
Our investigation of the halo models begins with isothermal and two-temperature models, which we
assume are either in CIE or in NEI. The results of fitting the model in which the halo is assumed to have
a single temperature (the isothermal model or $1T$ model) to our \Suzaku+\ROSAT\ data are shown
in the first row of Table~\ref{tab:2T model}. The noted error bars delineate the 90\% confidence
intervals. Hereafter, error bars calculated via XSPEC fits can be taken as 90\% confidence intervals.
The large $\chi^2$, however, indicates that the isothermal model may be too
simple for the hot gas in the Galactic halo along our sight line. Using the RASS data, \cite{kuntz00}
reached the same conclusion for the halo's hot gas in general. They found that the emission from the
halo's hot gas is best described by a two-temperature model (hereafter $2T$ model). Such a $2T$
model was also adopted in Paper I for the analysis of the \Suzaku+\ROSAT\ data, and provided good
fits to the spectra. A recent analysis of the \Chandra\ observation of a nearby edge-on galaxy
\citep[NGC~5775;][]{lijiangtao08} also shows that a $2T$ model is valid for the halo X-ray emission.
We tested the $2T$ model for the current project using the
\Suzaku+\ROSAT\ data, and display the results in the fourth row of Table~\ref{tab:2T model}. The $2T$
model provides a good fit to our \Suzaku+\ROSAT\ data. The temperatures we get
($\log T_1=6.12^{+0.02}_{-0.01}$ and $\log T_2=6.50^{+0.02}_{-0.02}$) are consistent with those
found by \cite{kuntz00} ($\log T_1=6.06^{+0.19}_{-0.20}$ and $\log T_2=6.46^{+0.12}_{-0.08}$).
However, the temperatures we get for the hot gas in our own Galactic halo are slightly lower than those
found by \cite{lijiangtao08} for NGC~5775 ($\log T_1\simeq 6.4$ and $\log T_2\simeq 6.8$). These high
temperatures may be connected with NGC~5775's  higher rate of star formation activity. To test the influence of
the uncertainties in the on- and off-filament column densities, we vary the on-filament $\nh$ between
$7.0\times10^{20}\pcms$ and $12.8\times10^{20}\pcms$ (corresponding to the range of values derived 
from the color excess measured by \cite{penprase98}). For the off-filament direction, we also test a 
column density as low as $0.5\times10^{20}\pcms$ and as high as $2.0\times10^{20}\pcms$. We find
that the temperature of the hot halo
component varies by up to 3\%, the temperature of the warm halo component varies by up to 20\%,
and the temperature of the LB component  varies by up to 45\% from those for the nominal \nh\ values.
The derived \ovii\ and \oviii\ intensities (see Section 5.1) vary by up to 22\% and 9\%, respectively. The
uncertainties in the
on- and off-filament X-ray absorbing column densities have little effect on the general conclusions made
about the halo's hot gas in this paper.

\subsection{Ionization State of the Halo's Hot Gas}
Here we examine the ionization state of the halo gas by comparing CIE and
NEI models for the halo emission. As discussed in the previous section, when the \ROSAT\ R12 are
included in the CIE fitting, we use different CIE models for the different datasets (i.e. RS code for 
\ROSAT\ and APEC for \Suzaku). It is, however, impossible to follow this procedure when using NEI
model(s) because a RS-code-based NEI model is not available in XSPEC. As a result, we fit our models
with NEI component(s) to the \Suzaku\ spectra only. The parameters of the LB component are fixed at the
values found from the previous CIE model fitting to the \Suzaku+\ROSAT\ data. This is because
the LB component is mainly constrained by the \ROSAT\  R12 data, which are not included in the fitting
here. We also fit corresponding CIE models to the \Suzaku\ spectra only (with the LB model
parameters fixed to the previously determined values; see the second row of Table~\ref{tab:2T model})
and use these results for comparison with those of the NEI modeling.

We first experiment with NEI modeling by testing an isothermal halo. We replace the CIE
($\mathtt{apec}$) halo component shown in row 2 of Table~\ref{tab:2T model} with an NEI
($\mathtt{nei}$) component. The NEI model has one more parameter $\tau=n_et$, where $n_e$ is the
electron density and $t$ is the time since the heating. CIE is reached when $\tau \gtrsim 10^{12}\uip$
\citep{masai94}. The fit results are displayed in row 3 of Table~\ref{tab:2T model}. The best isothermal
NEI halo model
is similar to the best isothermal CIE halo model in that their temperatures are similar, and with a
$\tau=35.0^{+15.0}_{-33.2}\times10^{12}\utau$, the NEI model is approximately in CIE.
It should be noted that the shape of the NEI halo component is somewhat
constrained by the fact that the LB component is fixed. However, the NEI model provides a better fit to
the \Suzaku\ spectra. While examining isothermal halo models may be informative, such models are
too simplistic. So we progress to two-temperature modeling.

In preparation for the $2T$ NEI modeling, we first establish a comparable CIE model, i.e. a model that
is fit to the \Suzaku\ data alone, but whose LB parameters are taken from the fit to the \Suzaku\ and
\ROSAT\ data. The parameters for this model are listed in row 5 of Table~\ref{tab:2T model} . We then
replace one of the CIE halo components with NEI component, while fixing the LB and the other halo
component parameters to the values listed on row 5. The resulting best fit models are listed on row 6
and 7 of Table~\ref{tab:2T model}. The low value of $\tau$ in the first halo component in row 6 suggests
that some of the halo gas may be in the process if ionizing. This is seen again and more strongly when
we allow both halo components to have NEI (see row 8).

\section{Constraining the EM Distribution of the Halo's Hot Gas using UV and X-ray Emission Lines}
\subsection{The \OVII\ and \OVIII\ X-ray Emission Lines}
The thermal properties of the hot gas in the Galactic halo can be constrained by emission line
measurements using the halo \ovi\ and \civ\ intensity measurements obtained from other sources and
our halo \ovii\ and \oviii\ measurements from the \Suzaku\ data. We will outline an EM
distribution that spans a temperature range of 2 dex, i.e. $T\sim10^{5.0}-10^{7.0}\K$.  In
Paper I, the \ovii\ and \oviii\ line intensities were measured for both the LB and the
Galactic halo. Here, focusing on the halo component, we measure the \ovii\ triplet ($\sim$570~eV) and
\oviii\ Ly$\alpha$ doublet ($\sim$650~eV) line intensities again, using a different method from Paper I.
We begin our measurements with $2T$ CIE model that was fit to the \Suzaku\ data, i.e. row 5 of
Table~\ref{tab:2T model}. This model provides a good fit to our spectra. Although later in this article we
show that the $2T$ CIE model is unable to explain the \ovi\ and \civ\ observations, the accuracy of the
measurement of the \ovii\ and \oviii\ intensities  is mainly determined by the goodness of the fitting to the
\Suzaku\ data rather than the physical meaning of the model.

The earlier $2T$ CIE modeling yielded the temperatures and EMs of the hot halo gas.
From these values we can calculate the intensity of the chosen emission line or complex, $I$, from
\begin{equation}
I = \frac{1}{4\pi} \frac{n_H}{n_e} \epsilon(T)\int n_e^2 dl,
\label{equ:erc}
\end{equation}
where $\int n_e^2 dl$ is the emission measure, $EM$, and $\epsilon(T)$ is the emission coefficient.
The factor of $n_H/n_e$ adjusts for the fact that the values of $\epsilon(T)$ tabulated in the APEC
database are normalized using $n_e n_H$ rather than $n_e^2$. We use the APEC emission coefficient
for consistency with our earlier fitting, in which we used the APEC model.

The APEC database lists line emissivities for a finite number of temperatures. In cases where the
temperature we are interested in is between two tabulated temperatures, we interpolate to obtain the
emissivity at our temperature of interest. To calculate the halo's intensity in the \ovii\ triplet, we use
Equation~(\ref{equ:erc}) and the temperatures and EMs obtained from the CIE model
fitting. We include the contributions from the resonance, forbidden, and intercombination lines. In the
case of the $2T$ CIE model we sum the contributions of the two halo components. We calculate the
\oviii\ emission line intensity in a similar way, including both components of the Ly$\alpha$ doublet.
The \ovii\ and \oviii\ line intensities obtained from fitting the $2T$ model to the \Suzaku\ spectra
are shown in the first row of Table~\ref{tab:O emission}.

We check our measurements of the \ovii\ and \oviii\ line intensities using the following independent
method. We add four $\delta$ functions to our $2T$ model to represent the \ovii\ and \oviii\ emission:
two unabsorbed $\delta$ functions represent the LB oxygen emission and two absorbed $\delta$ 
functions represent the halo's oxygen emission. We also ``turn off'' the oxygen line emission in the
APEC spectra for the halo components, and fit this new model to our \Suzaku\ data with all of the other
parameters fixed at their previously determined values. The energies of the two emission lines are also
free parameters to be determined by the fitting. The  best-fitting halo oxygen line intensities are shown
in the second row of Table~\ref{tab:O emission}. Note that the \ovii\ and \oviii\ intensities listed in
Table~\ref {tab:O emission} are intrinsic intensities; in effect, the observed intensities have been
deabsorbed with respect to the absorption due to intervening material along the sight line. A similar
technique was used in \cite{henley07} to measure the LB oxygen line intensities. In that case, the
oxygen emission from the LB APEC model was ``turned off" by setting the oxygen abundance to zero.
Here, we refine the technique used in \cite{henley07} slightly. Note that the database used by the APEC
model is made up of two files, $apec\_v1.3.1\_coco.fits$ and $apec\_v1.3.1\_line.fits$. The former is
used for calculating the continuum emission and the latter for the emission lines. Before running the
fitting procedure, we modified the emission line database by setting the emissivities for all of the \ovii\ 
and \oviii\ lines to zero but did not change the continuum database. As a result, best-fitting delta
functions account for only the oxygen line emission and not the oxygen continuum emission.
The various techniques for measuring the halo's oxygen line intensities give consistent results.
Also, as mentioned at the end of Section 4.1, the \ovii\ and \oviii\ intensities are not much affected by the
uncertainties in the X-ray absorbing column densities. Henceforth we shall use the \ovii\ and \oviii\
intensities obtained from our standard $2T$ model.

\subsection{The \OVI\ and \CIV\ UV Emission Lines}
Ultraviolet observations of \ovi\ and \civ\ resonance line doublet emission from the filament region are
also available. Using the \FUSE\ observations of the same directions \citep{shelton03}, we obtain a
de-absorbed intensity from the \ovi\ doublet ($\lambda\lambda 1032$, 1038) of
$7750^{+950}_{-1090}\lu$ (line unit; $\flux$) with 1$\sigma$ error bars, assuming an absorbing
$\nh$ of $1.9\times10^{20}\pcms$. We also have an off-filament \SPEAR\ observation of the
\civ\ resonance line doublet ($\lambda\lambda 1548$, 1550) to a region of size $\sim4'\times4'$
and centered at a direction ($l=279.7, b=-47.2$) less than $1.0\degr$ away from our off-filament line
of sight (J. Kregenow 2006, private communication). 
Since we have a \civ\ observation for only one direction, we cannot remove the LB contribution
as we do for the other lines using shadowing. However, as the LB is measured to have a temperature of
$\sim$$10^6\K$, it is not expected to emit much in the UV band. This has been confirmed by the \ovi\
observations which are sensitive to hot gas of temperature $\sim$$3\times10^5\K$: the $1\sigma$ upper
limit of the LB contribution to the \ovi\ doublet is only $\sim$$500\lu$, or less than 10\% of the emission
from the halo \citep{shelton03}. The ionization potential for \ciii$\rightarrow$\civ\ is lower than that of
\ov$\rightarrow$\ovi, and thus, collisionally ionized \civ\ is sensitive to gas with even lower temperatures.
The $\sim$$10^6\K$ Local Bubble will therefore contribute even
less to the \civ\ emission than it does to the \ovi\ emission, and so we attribute all of the \civ\ doublet
emission to the halo. The neutral hydrogen column density for the sight line of the \civ\ observation is
found from the LAB Survey map \citep{kalberla05} to be $\nhi=2.0\times10^{20}\pcms$.
Using the empirical relation $\nhi/E(B-V)=4.93\times10^{21}\pcms$ from \cite{diplas94} and
the extinction curve from \cite{fitzpatrick99}, we get a deabsorbed \civ\ doublet intensity of
$7780\pm2680\lu$ from the original observed value of $5790\pm2000\lu$. Since \civ\ emission may
also arise from photoionized media, we are going to take this value as an upper limit to constrain our
model around the temperature of $\sim10^5\K$. The measurements of the
four emission features (\civ, \ovi, \ovii, and \oviii) are summarized in Table~\ref{tab:4 emission lines}.

\subsection{EM Distribution Outlined by the Four Emission Lines}
Here, we calculate the halo's EM as a function of temperature from the intrinsic \civ, \ovi,
\ovii\, and \oviii\ intensities, Equation~(\ref{equ:erc}), and the theoretical emission coefficient. In order to
maintain consistency with our \Suzaku+\ROSAT\ modeling, in which the RS database \citep{raymond91}
was used for photon energies $\la0.3\kev$ and the APEC v1.3.1 database \citep{smith01a} was used
for photon energies $\ga0.3\kev$, we take the emission coefficients for the \civ\ and \ovi\ resonance line
doublet from the RS database and the coefficients for the \ovii\ triplet and \oviii\ Ly$\alpha$ line from the
APEC database. The upper panel of Figure~\ref{fig:4 emission lines} shows the emission coefficients
as a function of temperature. Because each emission coefficient covers a finite temperature range, we
cannot simply
assign all of the corresponding emission to a single temperature. Instead, for each ion, we determine the
average emission coefficient for the temperature range for which the emission coefficient exceeds 10\%
of its peak value, $\bar{\epsilon}$; we multiply this by the range of $\log T$, i.e. $\Delta\log T$, then
take $4\pi~n_e/n_H~\bar{\epsilon}~\Delta\log T$ to find the average EM per unit $\log T$.
The results for the four ions, plotted in the lower panel of Figure~\ref{fig:4 emission lines}, trace out
the halo's DEM function. The circles mark the temperature at which the
emission coefficients peak, the horizontal bars mark the temperature range over which the emission
coefficient exceeds 10\% of its peak value, and the vertical bars are error bars calculated from the
errors on the intensities.

\section{DEM Models of the Halo's Hot Gas}
The EM distribution or DEM function outlined by the four ions in
Figure~\ref{fig:4 emission lines} show that the hot gas in the Galactic halo is not isothermal.
The X-ray data also disallow an isothermal halo. But, the X-ray data do not disallow a $2T$ halo.
However, as shown in \cite{henley07}, a $2T$ model fit to the \xmm\ data for our directions
significantly under-predicted (by $3.3\sigma$) the \ovi\ intensity measured by \FUSE. Although the
\xmm\ observations are contaminated by SWCX emission, when we repeat the \ovi\ prediction using
our \Suzaku+\ROSAT\ fit results, we find that the discrepancy between the modeled and observed
intensities is even larger. Our $2T$ model fit to the \Suzaku+\ROSAT\ data predicts an intrinsic
\ovi\ intensity of $160\pm27\lu$, which is about 50 times smaller than the intrinsic intensity,
$7750^{+950}_{-1090}\lu$, which was calculated from the \ovi\ intensity observed by \FUSE\
\citep{shelton07} for an assumed $N_H$ of $1.9\times10^{20}\pcms$.

The fact that the isothermal and $2T$ models are inconsistent with X-ray and UV measurements
implies that the isothermal and $2T$ models may be over simplified. More sophisticated models have
been proposed by different authors using observations that probe a relatively wide temperature range.
Based on \ovi, \ovii, and \oviii\ absorption line measurements, \cite{yao07a} discussed the
non-isothermality of the hot gas in the Galactic halo and proposed a power-law DEM model. Assuming
exponential temperature and density distributions with respect to the height above the Galactic disk,
they were able to determine the slope and temperature range of the power law. Another power-law
DEM model covering a temperature range of $\sim$$10^{5.0}-10^{6.5}\K$ has been constructed by
\cite{shelton07} for the Galactic hot gas based on \FUSE\ \ovi\  and \ROSAT\ R12 ($1/4\kev$) and R67
($1.5\kev$) observations for our on-filament and off-filament directions. As these DEM models were
proved successful for hot halo gas within certain temperature ranges, we test various DEM models
to see if they are consistent with our set of observations covering a wide range of temperature as shown
in Figure~\ref{fig:4 emission lines}.

\subsection{Power-Law DEM Models}
\cite{shelton07} took the differential path length to be a power-law function of temperature
$dl=BT^\beta d\ln T$ for $T_1<T<T_2$. For an isobaric gas, this results in a power-law DEM model
of the form
\begin{equation}
	\frac{d\textnormal{EM}(T)}{d\log T} \propto
	\begin{cases}
		\left( \frac{T}{T_2} \right )^{\alpha} & \text{if $T_1 < T < T_2$,}\\
		0	& \text{otherwise,}\\
	\end{cases}
\label{equ:power law}
\end{equation}
where $\alpha=\beta-2$. \cite{shelton07} fixed the low-temperature cutoff at $T_1=10^{5.0}\K$ because
gas of lower temperature makes negligible contributions to the \ovi\ intensity. For
$N_H=2.0\times10^{20}\pcms$ they obtained $\alpha=-0.05\pm0.17$ and $T_2=10^{6.4}\K$.

The model is also shown in the lower panel of Figure~\ref{fig:DEM model} as the dotted line. We can
see a good agreement between the model and our \ovi\ emission line data point. The agreement is
expected, because both the model and the \ovi\ data point are produced using the same \ovi\
measurements (the small discrepancy is mainly due to the difference in the oxygen abundances,
emission coefficients and absorption column density assumed in \cite{shelton07} and this paper; see
below for more details).

In this section we would like to rework the \citeauthor{shelton07} model, to see if it is also consistent with
the \Suzaku\ observations which cover a higher energy band. But before doing this, we would like to
take  the opportunity to improve upon some of their approximations. For example, they used very
modern abundances in calculating the \ovi\ intensity (i.e. $O/H=4.57\times10^{-4}$ from \cite{asplund04}),
while the RS code they used for the modeling of the \ROSAT\ spectra relied upon an older set of
abundances from \cite{anders89}. Here we adopt a single set of modern and consistent abundances for
the modeling of all of the data. Following the argument in Paper I, the abundance table from \cite{wilms00}
 is used in this paper with an oxygen abundance of $O/H=4.90\times10^{-4}$. In
order to predict the \ovi\ and R12 intensities for comparison with the measurements, we need
the \ovi\ emission rate coefficient, $\rovi(T)$, and the R12 emission rate coefficient, $r_{12}(T)$, which
are equivalent to $1/4\pi~n_H/n_e~\epsilon(T)$ in Equation~(\ref{equ:erc}). We extract these from the
RS code because both the \ovi\ emission line and the R12 band are at energies lower than $0.3\kev$.
The \ovi\ emission rate coefficient, $\rovi(T)$, is consistent with the $\epsilon(T)$ used in the previous
section for the evaluation of the \ovi\ doublet intensity. Our R12 emission rate coefficient, $r_{12}(T)$,
is obtained by convolving the \ROSAT\ R12 response function with spectra of various temperature
plasmas that were calculated using the RS code. Our R12 emission coefficient, $r_{12}(T)$,
pertains to the intrinsic intensity and we use the deabsorbed R12 count rate in our calculation. In contrast,
\cite{shelton07} used an emission coefficient, $r_{12}(\nh,T)$, that accounted for the absorption
of material along the sight line and used the absorbed halo R12 count rate in their calculation of the
DEM. Our $r_{12}(T)$ is equivalent to their R12 emission coefficient, $r_{12}(\nh,T)$, when
$\nh$ is set to zero.

To measure the deabsorbed R12 intensity from the Galactic halo's hot gas, we fit our $2T$ CIE
model simultaneously  to the on- and off-filament \ROSAT\ spectra extracted from the RASS database
\citep{snowden97}. Here, the RS model is used for all of the LB and halo components because we
focus on the R12 band. Throughout this paper, the off-filament hydrogen column density
is taken to be $\nh=1.9\times10^{20}\pcms$ except for the testing of the influence of the uncertainty
in this value, so readers are reminded to compare our results with
those obtained by \cite{shelton07} for the most similar $\nh$ value ($2.0\times10^{20}\pcms$). The
intrinsic R12 emission from the Galactic halo's hot gas is then easily calculated using the fit results. Our
best-fitting spectrum yields R12$=(3740\pm450)\times10^{-6}\countrate$. Taking the deabsorbed
\ovi\ intensity as $I_{\ovi}=7750^{+950}_{-1090}\lu$, yields
$I_{\ovi}/R12=2.07^{+0.36}_{-0.38}\times 10^6\frac{\lu}{\countrate}$. Following \cite{shelton07}, we
determine the index of the power-law model using their Equation~(8) which we reproduce here:
\begin{equation}
\frac{\Iovi}{R12}=\frac{\int^{\ln T_{2}}_{\ln T_{1}}\rovi(T)T^{\alpha}d\ln T}
                                      {\int^{\ln T_{2}}_{\ln T_{1}}r_{12}(T)T^{\alpha}d\ln T}.
\label{equ:iovi/r12}
\end{equation}
However, we replace their R12 emission rate coefficient, $\rovi(\nh,T)$, with the deabsorbed 
emission coefficient $\rovi(T)$ so as to be consistent with our deabsorbed measurement of the R12
count rate, and we use $\alpha=\beta-2$. Following \cite{shelton07}, we set the low-temperature cutoff,
$T_1$ to be $10^{5.0}\K$. We then test this model for various high-temperature cutoffs to see if the
model is also consistent with our \Suzaku+\ROSAT\ data. For each high-temperature cutoff, $T_2$, we
calculate the slope $\alpha$ using Equation~(\ref{equ:iovi/r12}). Since both the value of \Iovi\ and R12
(not just the ratio) are known to us, we can also determine the constant of proportionality in
Equation~(\ref{equ:power law}). With this constant, the model predictions for the \ovii\ and \oviii\ 
line intensities are then calculated using the analogs of the numerator of Equation~(\ref{equ:iovi/r12}).
The \rovii\ and \roviii\ emission rate coefficients are extracted from the APEC database as these two
lines have photon energies $>0.3\kev$.

Here, our technique diverges from that of \citeauthor {shelton07}. We try several plausible values
for the high-temperature cutoff ($\log T_2$=6.06, 6.24, 6.36, and 6.54). For each value, we
determine $\alpha$ from Equation~(\ref{equ:iovi/r12}), determine the constant of proportionality for
Equation~(\ref{equ:power law}), and then calculate the \ovii\ and \oviii\ line intensities predicted by the
power-law model. Table~\ref{tab:SPL model 1} lists the $\alpha$ and the \ovii\ and \oviii\ intensities
for each examined $T_2$ values. The observationally determined intrinsic \ovii\ intensity
($9.98^{+1.10}_{-1.99}\lu$) is best modeled by the second case, that having $T_2=10^{6.24}\K$,
$\alpha=0.54$, and $\Iovii=10.4\lu$. This case slightly overpredicts the \ovii\ intensity but is within the
observational error bars. The other cases over- or underpredict the intensity by $>50\%$. However,
the model significantly under-predicts the \oviii\ intensity 
 ($0.49$ vs. $2.66^{+0.37}_{-0.30}\lu$).
If we increase $T_2$ to $10^{6.36}\K$ to improve the agreement between the model predicted \oviii\
intensity ($2.97\lu$) and the observed intensity, the model then more severely over-predicts the \ovii\
intensity ($18.4\lu$). The poor correspondence between model and observation can be seen more
directly in the upper panel of Figure~\ref{fig:DEM model}, where the power-law model with
$T_2=10^{6.54}\K$ is shown as the dotted line, together with the four emission
line data points. Since the model is derived from  the \FUSE\ \ovi\ observations, it does match the \ovi\
emission line data point well. The \ovii\ and \oviii\ line data points, however, obviously drop away from
the model. They and the \ovi\ data cannot simultaneously be explained by a single power-law model
with any choice of the high-temperature cutoff, as demonstrated in Table~\ref{tab:SPL model 1}. 

The power-law DEM model of \cite{shelton07} was constructed mainly based on the \FUSE\ \ovi\ and
\ROSAT\ R12 measurements, but we have found that it is impossible to extend the model to a higher
temperature and make it consistent with the \ovii\ and \oviii\ measurements. We now test the
power-law model in another way. We first fit a power-law model to the \Suzaku+\ROSAT\ data and then
determine if it is consistent with the \civ\ and \ovi\  measurements. The exponent, $\alpha$, and the
high-temperature cutoff, $T_2$, are free parameters of the model to be determined by the fitting. The
low-temperature cutoff, $T_1$, is not well determined by the fitting and is therefore fixed. We test
models with various  choices of $T_1$ and present the results in Table~\ref{tab:SPL model 2}.
While this power-law model fits the \Suzaku+\ROSAT\ data quite well for all of the choices of the
low-temperature cutoff, $T_1$, none of the values of $T_1$ results in predicted \civ\ and \ovi\
intensities that are both consistent with the measurements. For a low-temperature cutoff of
$T_1=10^{5.24}\K$ the predicted intrinsic \civ\ intensity is consistent with the observational value, but
the predicted intrinsic \ovi\ intensity is far too large ($46,500$ vs. $7750^{+950}_{-1090}\lu$), whereas
for $T_1=10^{5.57}\K$, even though the predicted \ovi\ intensity is consistent with the observational
value, the predicted \civ\ intensity is too small ($181\lu$) unless essentially all of the
observed \civ\ emission ($7780\pm2680\lu$) is due to photoionized gas in high-pressure photoionized
regions around hot stars. When the low-temperature
cutoff is $T_1=10^{5.76}\K$, both the \ovi\ and \civ\ intensities predicted by the model fall below the
observed values. Again, this is shown directly in the upper panel of Figure~\ref{fig:DEM model}. The
dot-dashed line, which represents the power-law model that best fits the \Suzaku+\ROSAT\ data and
has a low-temperature cutoff $T_1=10^{5.76}\K$, is consistent with the \ovii\ and \oviii\ emission line
data points, but is not consistent with the \ovi\ and \civ\ data points in the lower temperature range.

\subsection{A Broken Power-Law DEM Model}
Motivated by the partial successes of our power-law model that was fit to the \Suzaku+\ROSAT\ data
and the power-law model that was found for the \ovi\ and $1/4\kev$ emission
\citep[patterned after][]{shelton07}, we investigate a broken power-law DEM model of the form
\begin{equation}
	\frac{d\textnormal{EM}(T)}{d\log T} \propto
	\begin{cases}
		\left( \frac{T}{T_2} \right )^{\alpha_1}	& \text{if $T_1 < T < T_2$,}\\
		\left( \frac{T}{T_3} \right )^{\alpha_2}	& \text{if $T_2 < T < T_3$.}\\
	\end{cases}
\label{equ:BPL}
\end{equation}

Like the power-law model of \cite{shelton07}, the slope of the low-temperature ($T_1 < T < T_2$)
portion of the broken power-law DEM, $\alpha_1$, is constrained by the \FUSE\ \ovi\ intensity and
some of the \ROSAT\ R12 count rate, using Equation~(\ref{equ:iovi/r12}). We fix $T_1$ to the
value of $10^{4.8}\K$ (rather than $10^{5.0}\K$), so that the low-temperature portion of the model fully
covers the temperature regime probed by the \civ\ emission line observation. The slope of the
high-temperature ($T_2 < T < T_3$) portion, $\alpha_2$, is constrained by fitting to the
\Suzaku+\ROSAT\ data. While for a reasonable break temperature ($T_2 > 10^{5.5}\K$)
nearly all of the \ovi\ emission is
produced by the low-temperature portion of the DEM, the R12 emission is produced by both
portions. Hence, when using Equation~(\ref{equ:iovi/r12}) to calculate $\alpha_1$, the denominator on
the left-hand side should not be the total R12 count rate, but instead the fraction of the total R12
count rate produced by the low-temperature portion. We calculate this fraction (and ultimately the shape
of the broken power-law DEM) using the following iterative procedure. Because we do not know the break
temperature, $T_2$, a priori, we repeat the procedure using several different values of $T_2$ between
$10^{5.76}\K$ and $10^{6.37}\K$. We first fit the broken power-law DEM model to the \Suzaku+\ROSAT\
data with $\alpha_1$ fixed at some initial estimate, and $\alpha_2$ and the normalizations free to
vary. From these fit results we calculate the R12 count rates from the two portions of the DEM, and insert
the R12 count rate due to the low-temperature portion into Equation~(\ref{equ:iovi/r12}) in order to
calculate a new estimate of $\alpha_1$. We then re-fit our model to the \Suzaku+\ROSAT spectra with
$\alpha_1$ fixed at the new value, and with $\alpha_2$ and the normalizations free to vary. We repeat
this  procedure until the new R12 count rate due to the low-temperature portion of the broken power-law
model differs from the old one by less than 10\%. This method turns out to be quite efficient and stable.
The slopes converge within 4 or 5 iterations for a wide range of initial values of $\alpha_1$.

For the \Suzaku+\ROSAT\ data, the fit results with the broken power-law DEM model are summarized
in Table~\ref{tab:BPL model}. The break temperature of the model is well constrained, as it is easy to
understand that too high or too low a break temperature will make the broken power-law model
essentially fail for the same reason that the single power-law models failed. Various break temperatures
around $10^{6.0}\K$ are
tested. Besides the $\chi^2/dof$ value, the model predicted \civ\ intensity is used as a second constraint.
Of the models we tried, the one with a break temperature of $10^{6.02}\K$ and an $\alpha_2$ of -2.21
is preferred. This model has the smallest $\chi^2/dof$ as well as the best agreement with the observed
\civ\ intensity. 

We also test the effect of the uncertainties in the on- and off-filament X-ray absorbing column densities
on our BPL model. Using the same uncertainty ranges as those used in Section 4.1, we find that the \ovii\ and
\oviii\ intensities differ by 9\% and 8\% at most, respectively, and the slopes of the high-temperature
component (HTC) we get are consistently equal to $-2.2$. The uncertainty in the off-filament column density
does affect the slope of the low-temperature component (LTC) of the BPL and the $1/4\kev$ intensity predicted
by the model could vary by up to $\sim$20\%. Both of these effects, however, are not significant enough
to affect the main conclusions of this paper.


\section{Discussion}
\subsection{Possible X-ray Contamination in Our \Suzaku\ Spectra}
The assumption that our \Suzaku\ observations are not severely contaminated by SWCX X-rays was
supported in Paper I, where it was shown that the foreground oxygen intensities measured from the
\Suzaku\ spectra are consistent with zero. In addition, the on-filament and off-filament observations
were completed within a couple of days of each other during a minimum in the solar activity cycle.

Note that what we have called the LB component in our models really accounts for all of the foreground
emission, including the emission from the LB and that from SWCX, if there is any. Thus the \ovii\ and
\oviii\ intensities of our LB component (found by fitting our composite model to the data) actually
provide upper limits on the corresponding intensities due to SWCX. As shown in Tables~\ref{tab:2T model},
\ref{tab:SPL model 2}, and \ref{tab:BPL model}, all of the models (except for the $1T$ model, which we
have already shown does not provide a good fit to the spectra) yield similar results for the so-called
LB component, with $T\sim10^{5.95}\K$ and $EM\sim7\times10^{-3}\uem$. These fit results for the
LB component then predict foreground \ovii\ and \oviii\ intensities of  $\sim0.16$ and $\sim0.06\lu$,
consistent with the results found in Paper I using a different method ($1.1^{+1.1}_{-1.4}$ and
$1.0\pm1.1\lu$ for \ovii\ and \oviii\ respectively). Also, the foreground \ovii\ and \oviii\ intensities
predicted by our LB component are much smaller than the corresponding intensities due to the hot
gas in Galactic halo ($\sim10.0^{+2.7}_{-1.2}$ and  $\sim2.7^{+1.2}_{-0.3}\lu$ for \ovii\ and \oviii\
respectively). As a result, even if the small foreground \ovii\ and \oviii\ intensities found by our modeling
are all due to the SWCX, they should not affect our analysis of the hot gas in Galactic halo.

We now discuss the possible X-ray contamination from the stellar population. In processing the raw
\Suzaku\ data, we removed sources with $0.2-4.5\kev$ fluxes above a critical flux
$f_c=5\times10^{-14}\ergflux$. Unresolved X-ray emission from fainter stellar sources with fluxes
$<5\times10^{-14}\ergflux$, however, could be mixed with the diffuse X-ray emission from hot gas.
We therefore estimate the fraction of the observed X-ray emission that is due to stellar sources with
fluxes less than the critical flux $f_\mathrm{c}$. To do this, we used the X-ray luminosity function, for the
entire stellar population in the solar neighborhood, given in \cite{sazonov06}. To be conservative, we
use the space density profile for stars with $M_v>3.5$ from \cite{ojha96}. Assuming that the luminosity
function is independent of height, the total number of stars within our \Suzaku\ field of view is
\begin{equation}
N=\int n(z,L)dVdL=\int n(z)\Phi(L)\frac{z^2\Omega}{\sin^3|b|}dLdz,
\end{equation}
where $n(z)$ is the density profile as a function of height $z$, and $\Phi(L)=dN/dL_x$ is the X-ray
luminosity function, $\Omega=17.8'\times17.8'$ is the field of view of our \Suzaku\ observations, and
$b\simeq-45\degr$ is the Galactic latitude of our observations. Since the flux associated with each star is
$f=\frac{L}{4\pi d^2}=\frac{L\sin^2|b|}{4\pi z^2}$, the total X-ray flux from all of the stellar sources within
the field of view is
\begin{equation}
f_\mathrm{t}=\int\frac{L}{4\pi d^2}n(z,L)dVdL=\int_{z=0}^{z=\infty}\int_{L=0}^{L=\infty}\frac{1}{4\pi}n(z)\Phi(L)L
\frac{\Omega}{\sin|b|}dLdz.
\end{equation}
The total X-ray flux from fainter stellar sources with fluxes less than the critical flux $f_c$ is
\begin{equation}
f_\mathrm{s}=\int_{z=0}^{z=\infty}\int_{L=0}^{L=L_c}\frac{1}{4\pi}n(z)\Phi(L)L\sin|b|\Omega dLdz.
\end{equation}
where $L_c=4\pi f_\mathrm{c}[z/\sin|b|]^2$.

The space density profile for stars with $M_v>3.5$ from \cite{ojha96} is
\begin{equation}
n(z)\propto e^{-\frac{z}{h_1}}+0.074e^{-\frac{z}{h_2}},
\end{equation}
where the scale height of the thin disk is $h_1=260\pc$ and scale height of the thick disk is $h_2=760\pc$.
The X-ray luminosity function in the $0.1-2.4\kev$ band derived from the RASS for stars in the solar
neighborhood is shown in Figure~5 in \cite{sazonov06}. Following the form used fro the $2-10\kev$ X-ray
luminosity function given in the same paper, we parameterize the X-ray luminosity function for the $0.1-2.4\kev$
band as
\begin{equation}
\frac{dN}{d\log L_X}= K
	\begin{cases}
		(L_b/L_X)^{\alpha_1}, L_X< L_b\\
		(L_b/L_X)^{\alpha_2}, L_X> L_b.\\
	\end{cases}
\label{eq:dT/dt}
\end{equation}
From Figure~5 in \cite{sazonov06}, we estimate $L_b\simeq2.0\times10^{29}\erglu$,
$\alpha_1\simeq0.6$ and
$\alpha_2\simeq1.02$. We find that $\sim$80\% of the stellar X-ray emission was excluded by our
point source removal, and only $\sim20\%$ of the stellar X-ray emission remained to be mixed with
the diffuse X-ray emission from the hot gas. Note that fainter stars have a more extended distribution
than the brighter ones. Because we use the density profile for stars with $M_v>3.5$, we overestimate
the fraction of stars located at a larger distance from the plane and thus overestimate the fraction of
stellar X-ray emission that was not removed by our flux cutoff. Also, not taking into account X-ray
absorption makes the estimated fraction of stellar X-ray emission that is unresolved higher than the
true value, as absorption tends to increase with distance. Since the total X-ray emission from the
stellar population was estimated to be comparable with that from the hot gas in our Galaxy 
\citep{sazonov06}, and our conservative estimate shows that we have removed at least $\sim80\%$
of the stellar X-ray emission, we argue that our X-ray observations of the Galactic hot gas are not
badly contaminated by the X-ray emission from the stellar population.

\subsection{Comparing BPL DEM Models with $2T$ Models}
Using all of the \FUSE, \ROSAT\ and \Suzaku\ observations available for our sight lines, we have
successfully constructed a broken power-law DEM model covering a temperature range of
$10^{4.80}-10^{7.02}\K$. Before we discuss the physical implications of this model for the hot gas
in the Galactic halo, we point out that the $2T$ model, the single power-law model, and the
broken power-law DEM model all give similar values of $\chi^2$ when fit to the \Suzaku+\ROSAT\
spectra, although the $2T$ and single power law are inconsistent with the UV
observations. This means that the \Suzaku+\ROSAT\ data alone are insufficient to distinguish
between these various halo models. We can further demonstrate this fact with fake spectra generated
from one of our better-fitting broken power-law DEM and $2T$ models. We first generate fake on- and
off-filament \Suzaku\ and \ROSAT\ spectra from the broken power-law DEM model that has a break
temperature of $10^{6.06}\K$ (row 3 of Table~\ref{tab:BPL model}) with the XSPEC
$\mathtt{fakeit}$ command. When we fit the resulting spectra with our broken power-law DEM model,
we obtain $\chi^2/dof = 537.1/535$, whereas when we fit them with our $2T$ model we obtain
$\chi^2/dof = 538.9/535$. We also generate fake spectra from our $2T$ model. These spectra give
$\chi^2/dof = 564.7/535$ when fit with the $T_2=10^{6.06}\K$ 
broken power-law DEM model and $\chi^2/dof = 558.3/535$
when fit with the $2T$ model. The fact that both models give similar quality fits to both sets of fake
spectra shows that the \Suzaku\ spectra cannot distinguish between the broken power-law DEM
and $2T$ models. We thus conclude that even though the $2T$ model provides a good fit to the
\Suzaku+\ROSAT\ data, it may not necessarily be the real physical condition of the hot gas in the
Galactic halo. As for the single power-law model, which is plotted in the upper panel of 
Figure~\ref{fig:DEM model} as a dashed line, we can see that it mimics the high-temperature part of the
broken power-law model for the temperature range $\sim$$10^{5.8}-10^{6.5}\K$ covered by the
\Suzaku+\ROSAT\ data, but overproduces the \ovi\ intensity.

\subsection{Chemical Abundance Used in the Modeling}
Following \cite{henley07}, we have adopted the interstellar abundance table from \cite{wilms00} for
both emitting and absorbing gas in our spectral modeling. However, besides the abundance table from
\cite{wilms00}, five other abundance tables are available in XSPEC, namely those from \cite{anders89},
\cite{feldman92}, \cite{anders82}, \cite{grevesse98}, and \cite{lodders03}. Here we test the
$T_2=10^{6.06}\K$ BPL's sensitivity to the choice of abundance tables. For these tests, we first  chose
abundance tables which will be used for both the LB and halo components.  We then set the
low-temperature limit of the first part to be $T_1=10^{4.8}\K$ and the break temperature to be
$T_2=10^{6.06}\K$. We use the technique described in \S6.3 to determine the slope of the LTC. All of
the other parameters are determined by the fitting, which we repeat until the fit results stabilize. The
results are shown in Table~\ref{tab:abun BPL model}. The gross structure of the BPL remains almost
the same, regardless of the abundance table used. There are some differences in the slope of the
second part of the model, but given the error bars, the slopes are relatively consistent with each other.
We thus conclude that our model results are fairly independent of the choice of abundance table, and
our discussion of the properties of the hot gas in the Galactic halo based on the models is not affected
by the uncertainty in the abundances of the emitting and absorbing gas.

\subsection{Two-Component Scenario for the Halo's Hot Gas}
As clearly demonstrated by the spatial differences between the RASS maps in the R12, R45, and R67
bands, the $1/4\kev$ surface brightness is very patchy while the $3/4\kev$ and $1.5\kev$ maps are
much more uniform \citep{snowden97}. This difference between the $1/4\kev$ and $3/4\kev$ emission
cannot be ascribed entirely to the heavier absorption in the $1/4\kev$ band. Instead, the halo's
$1/4\kev$ and $3/4\kev$ X-rays may be produced by different components. The
hot gas which produces the majority of the halo's $1/4\kev$ surface brightness is not uniformly
distributed while the hotter gas that produces most of the $3/4\kev$ emission is more smoothly
distributed. A comprehensive discussion of a two-component model, of course, requires observations
of multiple sight lines, but here we would like to show that our study of one part of the sky is consistent
with a two-component model.

Our BPL model can be naturally divided into two parts, with each part being a power  law. We name
these two parts the LTC and the HTC in
accord with the temperature ranges they cover. The R12, R45, and R67 surface brightnesses made
by the different components of the BPL model are listed in Table~\ref{tab:BPL model prediction}. Before
we focus our attention on the R12 and R45 bands, we note that the total R67 count rate is dominated by
the contribution from the EPL component. This component is expected to be fairly isotropic across the
sky, in good agreement with the smoothness of the R67 band RASS map with the exception of  the
known extra radiation originating from the direction of the center of the Galaxy. The emission from the
HTC accounts for most of the total R45 count rate (i.e., $\sim55\%$), while the
EPL component accounts for the remainder ($\sim40\%$). The latter alone, however, cannot entirely
explain the smoothness of the RASS R45 map; therefore, we expect the halo's $3/4\kev$ X-ray emitting
gas (which is modeled in this paper as the HTC) to be fairly smooth. In the R12 band, the emission from
the LTC and the LB make a significant fraction ($\sim90\%$) of the total R12 count rate. The LTC
account for $\sim50\%$ of the non-local emission, which according to maps in \cite{snowden97}, has a
patchy distribution. Since the R12 count rate from the HTC should be fairly constant, then the LTC must
be responsible for the patchy appearance of the non-local R12 map in \cite{snowden97}. In a simplified
picture of the halo, our BPL accounts for the two-component nature of the halo, with the hot gas with a
patchy distribution being modeled by the LTC of our BPL and the hotter gas of much more uniform 
distribution being modeled by the HTC of our BPL. The possible origins of the two kinds of hot halo gas
will be discussed in Section 7.6.

\subsection{Comparison with the Other DEM Models for the Halo's Hot Gas}
DEM functions covering a similar temperature range have also been constructed by various
authors for the Galactic hot gas \citep[e.g.,][]{shelton07, yao07a, yao09}. As already mentioned in Section 6.1
and Section 6.2, our BPL model is actually motivated by the fact that the power-law model of \cite{shelton07}
cannot be simply extended to a higher temperature range and still be consistent with our \Suzaku\
measurements of the \ovii\ and \oviii\ intensities. For consistency with our \Suzaku\ observations,  a
HTC with a slope of $\sim$$-2$ must be added to a LTC. The result is a BPL model that is consistent
with all of the UV and X-ray observations. Our BPL model then mainly differs from the power-law model
of \cite{shelton07} by the addition of a HTC which covers a temperature range of
$10^{6.02}$-$10^{7.02}\K$, and accounts for the new \Suzaku\ observations which are sensitive to the
emission of the hot gas within that temperature range.

\cite{yao07a} also derived a power-law DEM function for the Galactic halo's hot gas, based on the
assumption that the $z$ distribution of both the temperature, $T$, and density, $n$, follow
exponential functions characterized by scale heights of $h_T$ and $h_n$. Setting $\gamma=h_T/h_n$,
their DEM function can be written as
\begin{equation}
\frac{d\textnormal{EM}(T)}{d\log (T)}\propto T^{2\gamma}.
\end{equation}
\cite{yao09} also used this functional form. Their power-law DEM models, constrained by the \ovi\ and
soft X-ray absorption line observations and diffuse soft X-ray emission observations for the hot gas in
the directions toward Mrk 421 \citep{yao07a} (dashed line) and LMC X-3 \citep{yao09}
(dot-dashed line), are also shown in the lower panel of Figure~\ref{fig:DEM model}.

Two significant differences between the power-law models of \cite{yao07a} and \cite{yao09} and
our BPL model can be noted. First, their power-law models have slopes of $1.2$ and $1.0$,
respectively. Not only are they inconsistent with our slope of $\sim$$-2.2$ for the HTC of our BPL, but
they are also much higher than the slope of $\sim$$0.3$ for the LTC. Also, their power-law models have
a much lower $d\textnormal{EM}(T)/d\log T$ value in the low-temperature range than our BPL model.
However, we have used an \ovi\ intensity measurement and \ROSAT\ R12 data to
constrain our model, while they did not include an \ovi\ intensity in their modeling. Their use of the \ovi\
absorption observations as a constraint ensures that their model is consistent with the \ovi\ column
density, but not necessary the \ovi\ intensity. In fact, \cite{yao09} did note that the \ovi\ intensity
predicted by their model is much less than typically observed for the halo. They suggested that most
of the observed \ovi\ intensity could be due to a second phenomenon, one that is not modeled by
their calculation: a transition temperature region between hot and cool gas. 

The values of $d\textnormal{EM}(T)/d\log T$ near $T=10^{6.3}\K$ in the models of \cite{yao07a} and
\cite{yao09} do agree with our BPL model in the high-temperature range, as both the power-law model
of \cite{yao09} and our BPL were constrained using the \ovii\ and \oviii\ intensity measurements in the
respective directions. Also, their use of $n(z)$ and $T(z)$ suggests that the modeled hot gas is smoothly
distributed (this is further supported by their similar results for
the two different directions). This is consistent with our interpretation of the $T>10^{6.0}\K$ gas,
as we also suggest that a portion of this gas component may be smoothly distributed. It would be
useful if additional data sets could be obtained, so that the analysis of the thermal the spatial properties
of the Galactic halo's hot gas could be repeated for a larger number of directions.

\subsection{Implications of Our Model to the Origin and Distribution of the Halo's Hot Gas}
It is widely believed that the hot gas in the halos of galaxies such as our own is due to either accretion
of the intergalactic medium (IGM) or stellar feedback. The thermal and spatial properties of the hot gas
resulting from these two mechanisms could be quite different. Therefore, some indications of the origin
of the hot gas may be obtained by comparing the thermal properties deduced from the observations
with those predicted by the theoretical models or simulations. In addition, given our two-component
scenario for the Galactic halo's hot gas, the morphology of each component may also give some useful
clues regarding the origin of the component.  In this section we consider the possible origin(s) of the
Galactic halo's hot gas by comparing our DEM model and emission line intensities with those predicted
by a simple accretion model and by SNR simulations. The morphology of the X-ray emission expected
from these two models is also compared with the RASS maps as another indicator of the origin of the
Galactic halo's hot gas.

\subsubsection{Accretion model}
Firstly, we consider a simple cooling model for IGM gas that is heated as it accretes onto the Galaxy.
For simplicity, we assume that the gas thermalizes as it falls though the Galaxy's gravitational potential,
comes to rest at some distance from the Galactic center, and then begins to cool radiatively.
Furthermore, if we ignore subsequent raising, falling, expansion, or contraction of the gas parcels, then
each parcel's potential energy can be taken to be constant because no additional work will be done on
the parcel. We estimate the temperature of the gas before it begins to cool radiatively, $T_0$,  to be
somewhat less than $10^{6.5}\K$, assuming that it was heated due to falling though the Galaxy's
gravitational potential to a Galacto-centric radius of $\sim$$8.5\kpc$ and assuming that the electrons
equilibrate with the stripped hydrogen and helium ions. Gas of this temperature contains \ovii\ and \oviii\
ions and emits 3/4 keV X-rays, whose spectrum is appropriate for comparison with observations.

The internal energy per unit volume of a parcel of gas at temperature $T$ is
$U_v = \frac{3}{2} n k T$, where $n$ is the number density of particles. The gas is assumed to be fully
ionized, so $n= n_e + n_i$, where $n_e$ and $n_i$ are the number densities of electrons and ions,
respectively. The gas is assumed to be fully ionized. The parcel of gas loses energy at a rate of
$dU_v / dt = - n_e n_i \Lambda(T)$, where $\Lambda (T)$ is the cooling function.
From the internal energy and the loss rate equation, we can determine that the plasma's temperature
changes at a rate of:
\begin{equation}
\frac{dT}{dt} = \frac{-2 n_e n_i \Lambda (T)} {3 n k}.
\label{eq:dT/dt}
\end{equation}

Equation~(\ref{eq:dT/dt}) can be used to determine the EM function of the hot gas that
accreted onto the Galaxy's halo if we assume that the gas accretes at a steady rate and then begins to
cool down from the same initial  temperature, $T_0$. We take the accretion rate, $dN/dt$, to be
constant, where $N$ is the number of accreted particles within a cross sectional area, $A$. Suppose
that the thickness of the accreted layer is $l$ and that the hot gas is produced and cools down in a way
that both the density, $n$ and the cross sectional area remain constant. This is the isochoric case. 
In this case, $dN = n A dl$ during the time interval $dt$. If the accretion proceeds for a time span that is
longer than the cooling time for the temperature regime of interest ($T\sim10^{4.8}-10^{6.5}\K$),
then the gas reaches a steady state with respect to temperature. Accreted gas enters the system at
a temperature of $T_0$, cools over time, and eventually leaves the temperature regime of interest.
However, as a given segment of gas is cooling, newer gas replaces it. Thus at any
given time, the quantity of gas of any given temperature (within the temperature regime of interest)
remains constant. The quantity of material within a temperature interval, $dT$, can therefore be
calculated from:
\begin{equation}
\frac{dN}{dT} = \left| \frac{dN}{dt} \frac{dt}{dT} \right| = \frac{dN}{dt} 
\frac{3 n k}{2 n_e n_i \Lambda(T)}.
\end{equation}

The EM associated with a given interval is $d EM = n_e^2 dl$, where $dl$ is simply
$dN / n A$. Thus,
\begin{equation}
\frac{d\textnormal{EM}(T)}{dT} = n_e^2 \frac{dl}{dT} =  \left( \frac{n_e}{n_i} \right)
\left( \frac{3 k}{2} \right) \left(\frac{1}{A}\frac{dN}{dt}\right) 
\frac{1}{\Lambda(T)}.
\label{eq:dEM/dt}
\end{equation}

This equation can easily be compared with our plotted DEM function given that $d\log T=\log(e)dT/T$,
$\Lambda(T)$ for isochoric, solar abundance, CIE gas is tabulated in \cite{sutherland93}, 
and noting that $n_e/n_i$, $3k/2$, and $(1/A)dN/dt$ are constants, although, admittedly, $(1/A)dN/dt$ is of
unknown value.

A scaled version of the DEM function for this simple steady state cooling scenario is plotted in
Figure~\ref{fig:cooling model}, together with our broken power-law model. The two curves bear little
resemblance to each other. Allowing the accreted gas to have subsolar abundances by using the
$\Lambda(T)$ curve for 0.1 solar metalicity CIE gas in \cite{sutherland93} would slightly change the
curve's slope between $T=10^{5.0}-10^{6.0}\K$, but as shown in Figure~\ref{fig:cooling model}, would
not bring the theoretical DEM into agreement with the observationally determined DEM. 

While the DEM functions predicted by the simple isochoric accretion model and the approximation of
CIE do not match the BPL DEM function we got from fitting both the UV and soft X-ray data of our sight
lines, we note that, within the temperature range $10^{5.0}-10^{6.5}\K$, the accretion model predicted
DEM functions mimic the power-law DEM  functions of \cite{yao07a} and \cite{yao09}, which successfully
explain the soft X-ray emission seen on their lines of sight. This suggests that even though the accretion
model fails to be a single explanation for all of the UV and soft X-ray emissive hot gas in the Galactic
halo, it still might be a phenomenological explanation for the soft X-ray emissive portion of the hot gas.
To test this idea, we fit the accretion model DEM to our \Suzaku\ data only. The technique mentioned in
Section 6.2 is used again to construct a tabulated accretion DEM model. This time, the spectra for 50 different temperatures obtained from the APEC database are weighted by the accretion model DEM
function (Equation~(\ref{eq:dEM/dt})). We calculate the accretion DEM spectra this way for a grid of
values of high-temperature cutoff with the low-temperature cutoff being set to be $10^{5.0}\K$. The
high-temperature cutoff and the scaling of the model are then two free parameters to be determined by
the fitting. Fitting the accretion model to the \Suzaku\ data results in a reasonably good fit, with
$\chi^2/dof=574.8/533$. The accretion rate can also be estimated from the best-fit normalization value.
Our result is $\dot{M}/A=(14/23)(m_\mathrm{H}/A)(dN/dt)=1.35\times10^{-3}\ifrate$. The factor $14/23$
comes from assuming that the accreting gas is fully ionized with H:He=10:1.  If intergalactic material falls
evenly onto our galaxy across the whole disk, then $A=2\pi R^2$, where $R=15\kpc$ is the radius of the
disk, and the factor of 2 is for the two sides of the disk. For the whole galaxy we then have
$\dot{M}=1.9\acrate$. Technically, this is an upper limit, because some of the X-rays observed by
\Suzaku\ may come from other sources.

\subsubsection{SNR model}
Next, we consider the possibility that the gas was heated by an explosive event, such as an SN
explosion. For this comparison, we use the results of SNR simulations from the
series of papers by \cite{shelton98,shelton99,shelton06}. The simulations employed a Lagrangian
mesh hydrocode with algorithms that model shock dynamics, NEI and
recombination, nonthermal pressure and thermal conduction. In \cite{shelton06}, modeled SNRs
located between 130 to 1800 pc above the Galactic midplane. The density of the ambient medium at
these heights was taken from \cite{ferriere98}. As we can see in Figure~\ref{fig:io6r1245}, both the \ovi\
and X-ray emission of an SNR are significant only when the SNR is younger than $\sim10^6\yr$. The
number of SNRs of age $\leqslant10^6\yr$ that reside on a typical sight line can be estimated from
typical radius ($\lesssim100\pc$) of the SNR and the SN explosion rate (Equation (10) in
\cite{shelton06}). Because the typical number is small ($\lesssim0.05$), the probability of encountering
two or more SNRs of age $\leqslant10^6\yr$ on a sight line is tiny and we thus only compare our
observationally results with those predicted for a single SNR.

The predicted \ovi/R12 ratio for an SNR residing at $z\sim1300\pc$ (ambient density of
$n_0$=0.01 atoms$\uden$) has already been compared with the \FUSE\ and \ROSAT\ observations for
our directions \cite[][although different units and conversions were used in that paper]{shelton07}, and
the conclusion drawn that the observed ratio best matched that of a remnant at an age before the SNR
formed a dense shell. Here we extend the work by comparing with a greater number of simulated SNRs
and extending the comparison to the $3/4\kev$ X-ray band. We examine remnants evolving in ambient
densities of $n_0$=0.2, 0.1, 0.05, 0.02, 0.01, and 0.005 atoms$\uden$, corresponding to heights of
$z$=190, 310, 480, 850, 1300, and 1800$\pc$, respectively, SN explosion energies of $E_0=0.5$
and $1.0\times10^{51}\ergs$, and ambient nonthermal pressures of $P_{nth}=1800$ and
$7200\K\uden$. The integrated \ovi, R12, and R45 intensities predicted by various model SNRs are
shown in Figure~\ref{fig:io6r1245} as curves of different colors and line types. Although our \FUSE\
and \Suzaku\ observations may cover only a small portion of an evolved SNR, the exact positions of
our sight lines relative to the possible SNR(s) are unknown. However, we find that the variation of the
intensities from one sight line to another is fairly small, and the integrated SNR intensities are thus used
in comparing the SNR models with our observations.
Our observationally derived intrinsic \ovi\ intensity, $1/4\kev$ count rate and $3/4\kev$ count rate
are $7750^{+950}_{-1090}\lu$ (from Section 5.2), $3740\pm450\times10^{-6}$ R12 $\countrate$
(from Section 6.1), and $92^{+27}_{-31}\times10^{-6}$ R45 $\countrate$ (from the prediction of our
BPL model). Interestingly, the observationally derived \ovi\ intensity and R12 and R45 count rates
simultaneously match the predictions for the SNR that has $n_0$=0.01 atoms$\uden$,
$E_0=1.0\times10^{51}\ergs$, and $P_{nth}=7200\K\uden$ (corresponding to the dashed blue curve in
Figure~\ref{fig:io6r1245}) at an age of $\sim10^5\yr$. Although the other models are able to match the
halo \ovi\ intensity and R12 count rate, they are not able to match all observations simultaneously.

The consistency between the observations and the predictions for a single SNR, however, is not what
we expected from the discussion in Section 6.3, where we argued that most of the homogeneously
distributed $3/4\kev$-emitting gas may have different source than the inhomogeneously distributed
$1/4\kev$-emitting gas. Like the $1/4\kev$-emitting gas, gas traced by \ovi\ ions is also inhomogeneously
distributed. Because the number of SNRs encountered on a sight line is small, the sporadic SNe should
produce a patchy distribution of hot gas. Therefore, it is logical to relate the LTC of our BPL model, which
produces almost all of the \ovi\ emission and the majority of the $1/4\kev$ emission, to the SNR model, and
to relate the HTC of our BPL model, which produces most of the $3/4\kev$ emission, to the more
uniformly distributed hot gas. For this reason, we now compare the \ovi, R12, and R45 intensities
derived from the LTC with the predictions of
the simulated SNRs. As shown in Figure~\ref{fig:io6r1245}, the intensities derived from
the LTC are best matched by the predictions of the SNRs with $n_0$=0.02 atoms$\uden$ at an
age of $\sim1.8\times10^5\yr$ (green curves). Although the simulated SNRs overpredicts the LTC's
R45 intensity, the predicted value is still smaller than the halo's total intrinsic R45 intensity
and thus leaves room for R45 emission produced by more smoothly distributed hot gas.

While both the \ovi\ and R12 portions of our halo emission observations can be explained by a single
SNR, the collection of \ovi\ $column$ $density$ and R12 count rate measurements for the high latitude
sky do not show a constant ratio between \ovi\ and soft X-rays \citep{savage03}. We note, however,
that the variation in attenuation from one line of sight to another is not taken into account in these
surveys nor is the contribution from the local region. Also, SNRs create both \ovi\
and soft X-rays, but the ratio of \ovi\ to R12 changes as the remnant evolves. For the simulations we
discussed, the ratio of \ovi\ intensity to R12 count rate varies by a factor of $>$100 with age. For these
reasons, SNRs should not be expected to result in a constant \ovi\ to R12 ratio on all line sights across
the high latitude sky.

We would like to end the discussion of SNRs as a possible component of the halo's hot gas with two
interesting points. First, based on the filament's \IRAS\ $12\um$ to $100\um$ intensity ratio, \IRAS\
$60\um$ to $100\um$ intensity ratio, and \caii\ kinematics, \cite{penprase98} determined that the
filament has been heated, probably by a shock, suggesting that the filament could be a very old
SNR. Second, we note that the halo R12 maps of \cite{snowden98} clearly show that the filament and
our off-filament observation overlap with a roughly circular region of angular
radius $\sim5\degr$ that has an elevated R12 count rate and an unusually low R2/R1 ratio.
These characteristics imply the existence of an unusual feature in this direction. For comparison, the
angular radius of the simulated SNR can also be estimated from the distance and the predicted radius
of the simulated SNR. We find that for the $1.8\times10^5\yr$ old SNR that has $n_0=0.02\uden$,
$E_0=1.0\times10^{51}\ergs$ and $P_{nth}=7200\K\uden$, the angular radius is $\sim4.0\degr$,
which is consistent with the angular size of the distribution in the R12 maps.

\section{Summary and Conclusions}
In this paper we analyze the \Suzaku\ spectra of the ISM obtained from observations pointing toward
and to the side of an absorbing filament at high southern Galactic latitude. We take a joint analysis of
these data and \FUSE\ and \ROSAT\ observations of the same sight lines, in order to constrain the
thermal and spatial properties of the hot gas in the Galactic halo. Our main findings are as follows:

1. \ovii\ and \oviii\ emission line features due to the Galactic halo's hot gas are firmly detected using our
\Suzaku\ shadowing observations. Their intrinsic intensities are $9.98^{+1.10}_{-1.99}$ and
$2.66^{+0.37}_{-0.30}\lu$, respectively. These observations, together with the \FUSE\ observations of
emission from \ovi\ in the Galactic halo along our off-filament line of sight (intrinsic \ovi\ intensity
$=7750^{+950}_{-1090}\lu$) and \SPEAR\ observations of emission from \civ\ along a direction less
than $1.0\degr$ away from our off-filament line of sight (intrinsic \civ\ intensity $=7780\pm2680\lu$),
sample hot gas with
temperatures ranging from $\sim10^{5.0}$ to $\sim10^{6.5}\K$. These observations indicate a
non-isothermal distribution of the hot gas in the Galactic halo, which is consistent with the finding of
earlier authors who modeled the halo with two thermal components \citep[e.g.][]{kuntz00}, and that of
\cite{yao07a} and \cite{shelton07} who modeled the halo with power-law DEM
functions.

2. We construct a DEM model for the halo's hot gas, in which
$d\textnormal{EM}(T)/d\log T$
follows a broken power-law function of $T$. This model is consistent with the \SPEAR, \FUSE, \ROSAT,
and \Suzaku\ observations. The LTC of the broken power law covers the
temperature range from $10^{5.0}\K$ to $10^{6.02}\K$, has an index of $\alpha_1=0.30$  and is
mainly constrained by the \FUSE\ \ovi\ intensity and \ROSAT\ R12 count rate. The low-temperature
cutoff of this component can be extended to $10^{4.8}\K$ and is consistent with the $SPEAR$ \civ\
intensity. The HTC covers the temperature range from
$10^{6.02}\K$ to $10^{7.02}\K$, has an index of $\alpha_2=-2.21$ and is mainly constrained by
the \Suzaku\ X-ray spectra. If we take the X-ray emission in our sight line to be representative of that
of the entire halo, then we can estimate the $0.2-2.0\kev$ soft X-ray luminosity of our Galaxy's halo
to be $3.0\times10^{39}\ergpers$. Considering the spatial differences between the $1/4$ and $3/4\kev$
RASS maps, we propose that the Galactic halo's hot gas is composed of two components. The higher
temperature, more uniformly distributed component is represented by the HTC of our BPL,
while the lower temperature, less uniformly distributed component is represented by the LTC of our BPL.
Confirmation of this hypothesis, of course, requires observations in more sight lines.

3. We compare our BPL DEM model with the power-law DEM models of \cite{yao07a} and \cite{yao09}
for the Galactic halo's hot gas toward the directions of Mrk 421 and LMC X-3, respectively. Their models
assumed an exponential disk scenario. Comparing their power-law models with our
broken power-law model, we find that the curves are inconsistent. We propose that the fundamental
reason for this inconsistency is that we included \ovi\ emission information in our analysis, while they
did not.

4. We compare our results with the following scenarios:

(a) A simple IGM accretion and cooling scenario. In this scenario, we assume that intergalactic gas is
accreted onto the Galactic halo at a constant rate. The hot gas then cools radiatively such that a line of
sight though the accretion layer samples a range of temperatures. We derive the DEM
function for the accretion layer, finding that its shape is inconsistent with the broken power-law
model derived from the observations (see Figure~\ref{fig:cooling model}). It under-predicts the UV
intensity relative to the X-ray intensity. However, with a high-temperature cutoff value of $10^{6.5}\K$,
the X-ray emission predicted by the accretion model is loosely consistent with our \Suzaku\ observations.
Attributing all of the \Suzaku-band X-ray emission to the accretion model yields an accretion rate of
$1.35\times10^{-3}\ifrate$ or $1.9\acrate$ for the whole galaxy. This is an upper limit because some of the
X-rays seen by \Suzaku\ may have come from other sources.

(b) A SNR scenario. We use existing simulations of SNRs evolving at various heights above the disk.
We find that the observed \ovi, $1/4\kev$, and $3/4\kev$ intensities match the predictions of a
$\sim$$10^5\yr$ old SNR located at a height of 1300 pc above the disk. The predicted angular size of
such a remnant is consistent with a bright spot on the \ROSAT\ $1/4\kev$ map in the direction of our
observations. Because we suspect that a more smoothly distributed source supplements the $3/4\kev$
intensity of sporadic explosive events, such as SNRs, we also consider dimmer SNRs.
A slightly older SNR located nearer to the galactic plane can make the \ovi\ and $1/4\kev$ photons and
some of the $3/4\kev$ photons while `leaving room for' a smoother source of $3/4\kev$ photons (see
Figure~\ref{fig:cooling model}). The IGM accreted onto our galaxy could be a possible origin of the
smoothly distributed component of the hot gas.

\acknowledgments
We thank J. Kregenow for providing us with the \SPEAR\ \civ\ intensity measurements for
sight lines near ours. We would like thank the referee for very helpful comments.
This project was supported by NASA through the Long Term Space Astrophysics
Program under grant NNG04GD78G, through the \Suzaku\ Guest Observer Program under grant
NNX07AB03G, and through the \xmm\ Guest Observer Program under grant NNG04GB08G. This
research has made use of data obtained from the \Suzaku\ satellite, a collaborative
mission between the space agencies of Japan (JAXA) and the USA (NASA).

\bibliography{references}

\clearpage
\begin{figure*}[htp]
\centering{\includegraphics[angle=270.0,width=1.0\textwidth]{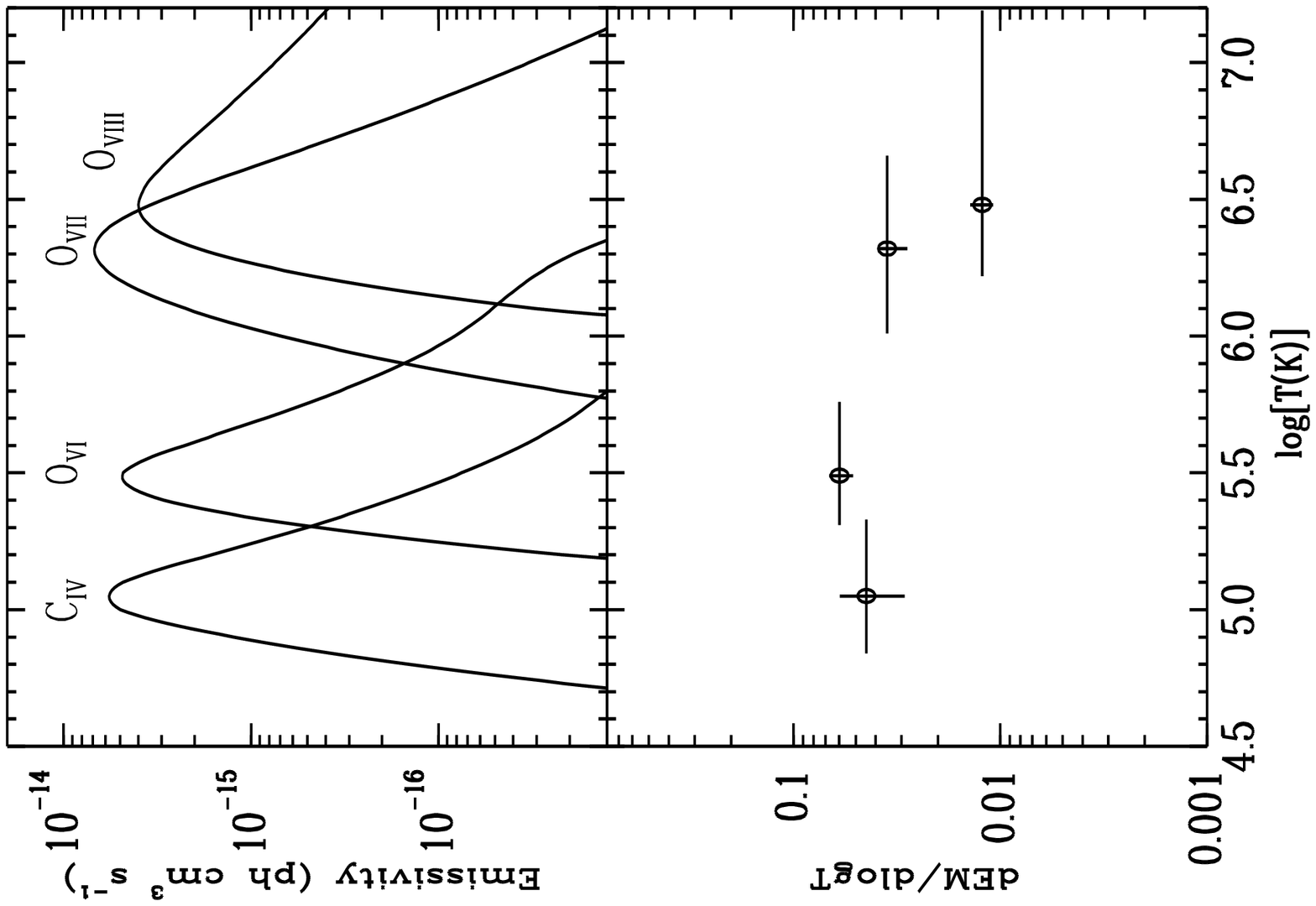}}
\centering\caption{Upper panel: The \civ\ ($\lambda\lambda1548, 1550$), \ovi\ ($\lambda\lambda1032,
1038$), \ovii\ (triplet at $\sim$$0.57\kev$), and \oviii\ ($\sim$$0.65\kev$) emission coefficients as a
function of the gas temperature. The values for \civ\ and \ovi\ have been scaled down by a factor of 1000
for clarity. The \civ\ and \ovi\ line coefficients are from the RS database, those for \ovii\ and \oviii\ are from
the APEC v1.3.1 database.  Lower panel: Galactic halo EM distribution as outlined by the four emission
lines. The four circles mark the temperature at which the emissivity peaks and the EM per unit $\log T$
needed to produce the halo's \civ, \ovi, \ovii, and \oviii\ intensities. The horizontal error bars show the
measurement of the intensities of the lines, and cover the temperature ranges for which the theoretical
emission coefficients are more than 1/10 of their peak values. The vertical error bars are derived from
errors on the intensities.}
\label{fig:4 emission lines}
\end{figure*}

\begin{figure*}[htp]
\centering{\includegraphics[angle=270,width=1.0\textwidth]{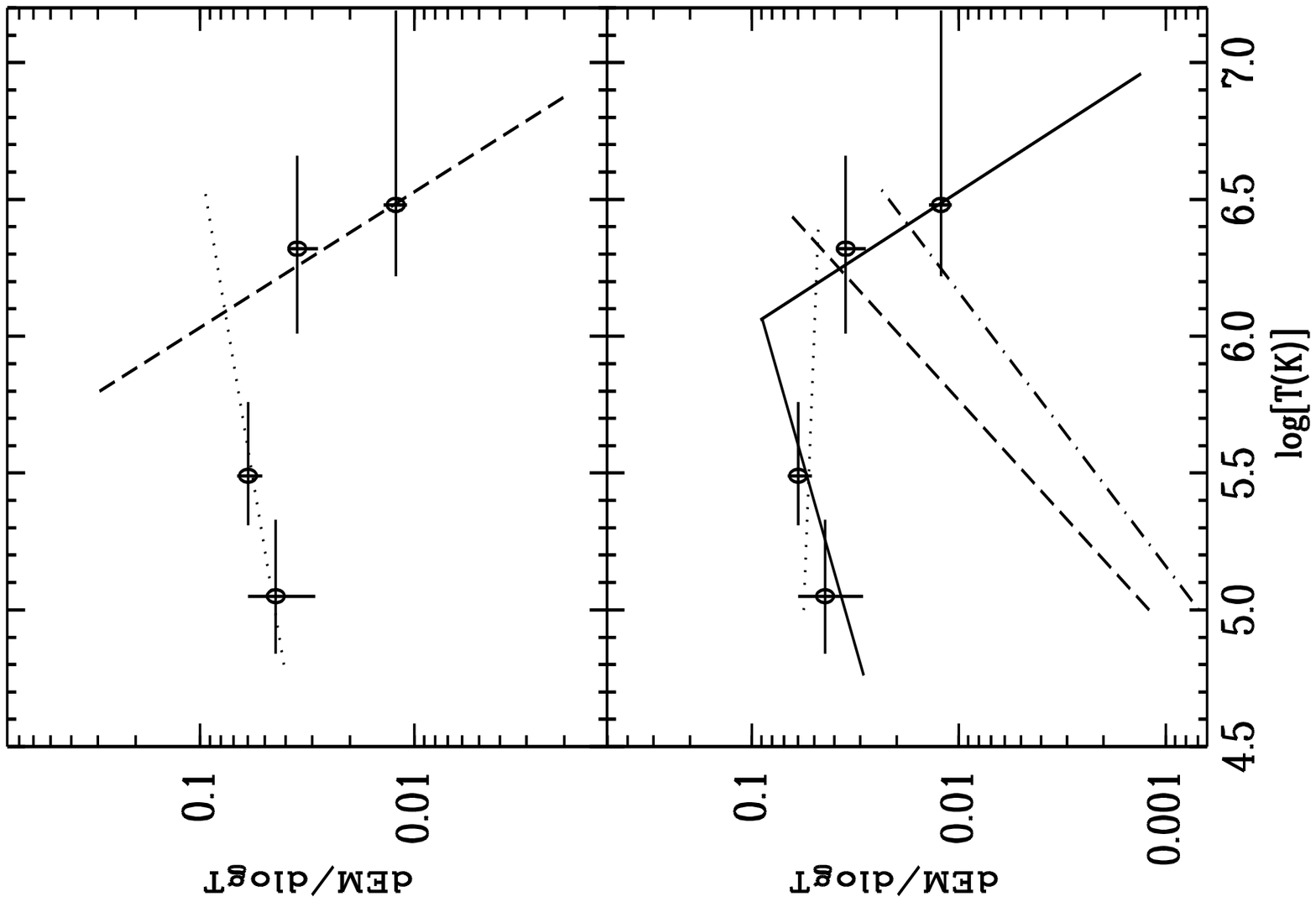}}
\centering\caption{Upper panel: Two power-law DEM models tested in this paper for the hot gas in the
Galactic halo: a power-law model similar to that of \cite{shelton07} (dotted line) and a power-law model
fitted to the \Suzaku+\ROSAT\ data only (dashed line). Lower panel: Our best-fitting broken power-law
DEM model (solid lines) in comparison with the power-law of \cite{shelton07} and the power-law model
of \cite{yao07a} towards the direction of Mrk 421 (dashed line) and that of \cite{yao09} towards the
direction of LMC X-3 (dot-dashed line).  (Note that we have reduced the DEM models of
\citeauthor{yao07a} and \citeauthor{yao09} using the oxygen abundance of \cite{wilms00}.) The \civ,
\ovi, \ovii, and \oviii\ data points from the lower panel of Figure \ref{fig:4 emission lines} are shown in
both panels for comparison.}
\label{fig:DEM model}
\end{figure*}

\begin{figure*}[htp]
\centering{\includegraphics[angle=270,width=1.0\textwidth]{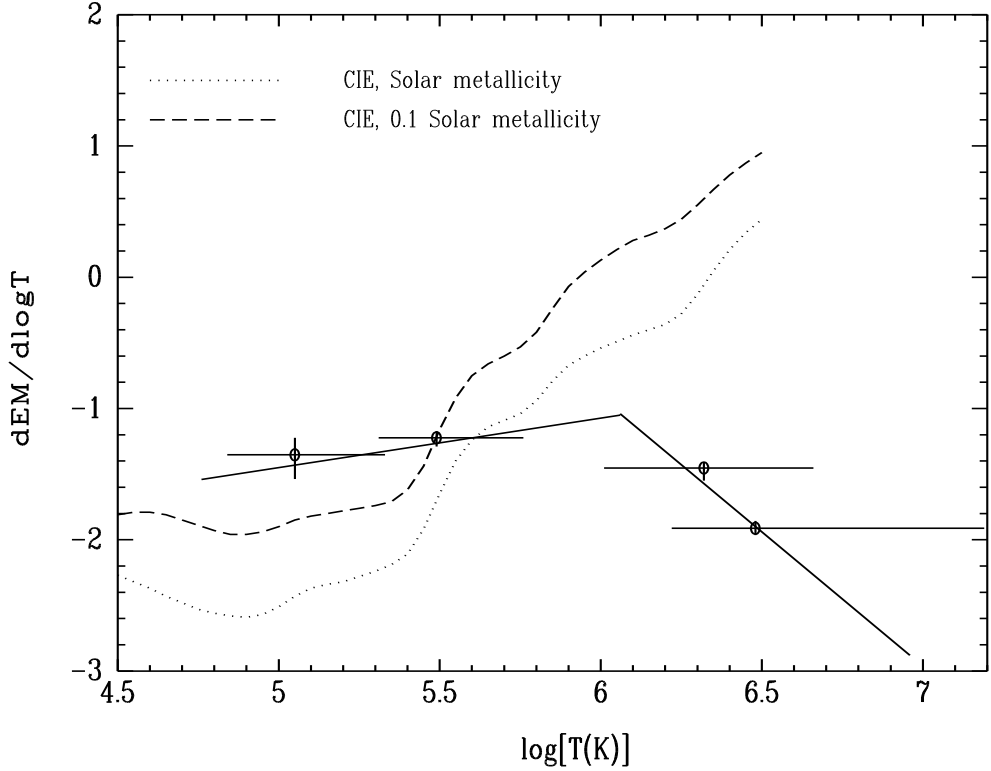}}
\centering\caption{Comparison between our BPL and theoretical cooling models:
Because the $(1/A)dN/dt$ coefficient is not known, we set it arbitrarily to a value that allows the accretion
DEMs to overlap our observationally determined BPL DEM. The scaled DEM functions predicted by
simple cooling models (dotted and dashed curves) do not resemble our BPL model above a
temperature of $10^{5.3}\K$, showing that our model is not consistent with a simple cooling picture, in
which the hot gas is first heated to an X-ray emitting temperature of $\sim10^{6.5}\K$ (see the text for
more details) and then cools radiatively. The dotted curve is the DEM function predicted by isochoric
cooling of solar metallicity CIE gas and the dashed one is for isochoric cooling of 1/10 solar metallicity
CIE gas.}
\label{fig:cooling model}
\end{figure*}

\begin{figure*}[htp]
\centering{\includegraphics[angle=270,width=1.0\textwidth]{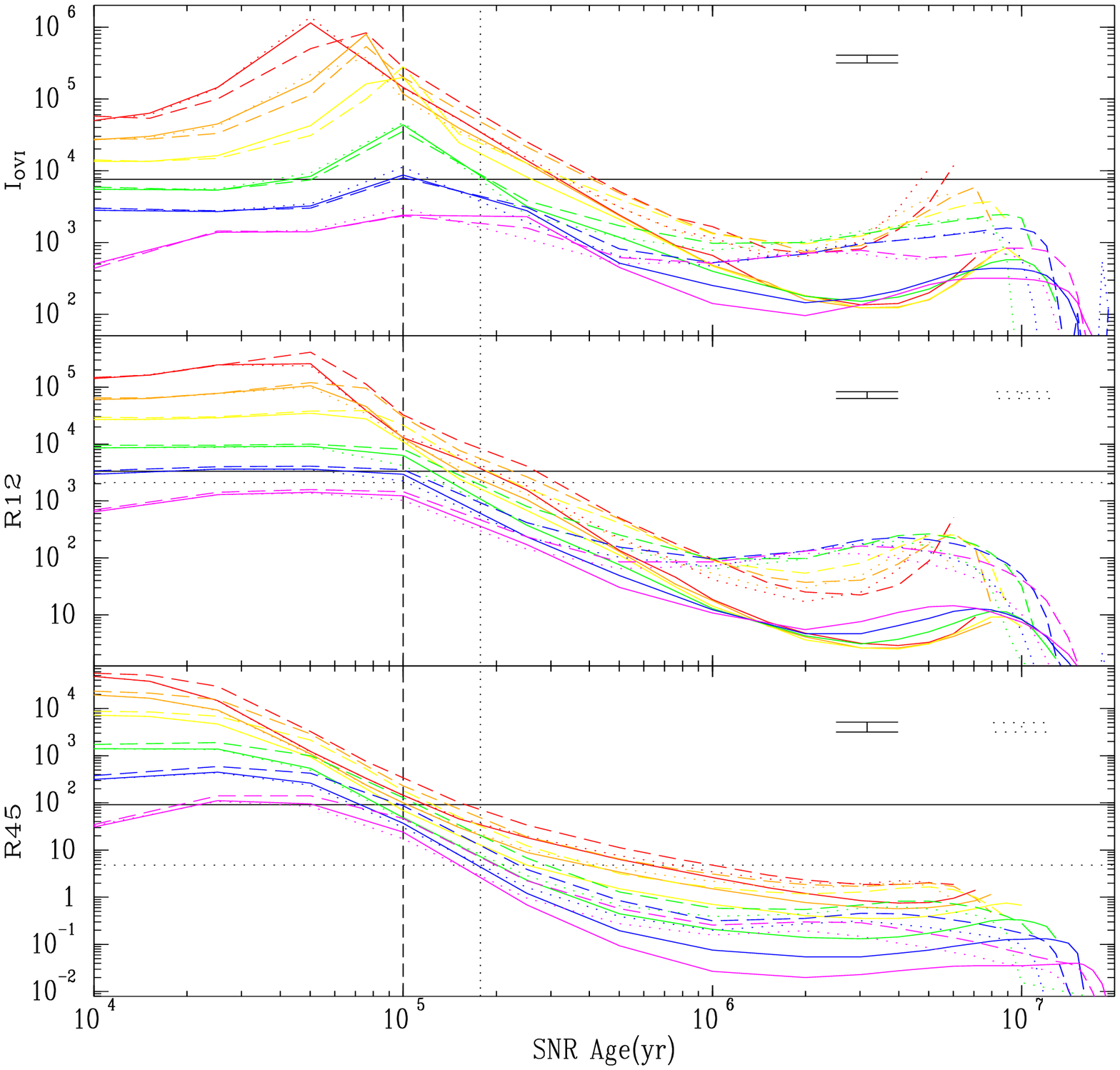}}
\centering\caption{Observationally derived halo \ovi, R12, and R45 intensities (horizontal solid
lines in each panel, with error bars noted by set of solid lines at the top right of the panel) are compared
with the predictions for various simulated SNRs (colored  curves). The \ovi\ intensities are in unit of
$\flux$ and the R12 and R45 intensities are in units of $10^{-6}\countrate$. The
ambient densities ($n_0$=0.2, 0.1, 0.05, 0.02, 0.01, and 0.005 atoms$\uden$) of the simulated
SNRs are keyed to the curve colors using a rainbow sequence (red, orange, yellow, green,
blue, and purple). The line types of the curves (solid, dotted, and dashed) distinguish the SN
explosion energy ($E_0$=0.5, 0.5, and $1.0\times10^{51}\ergs$) and ambient nonthermal pressure
($P_{nth}$=1800, 7200, and 7200$\K\uden$). The total halo \ovi, R12 and R45 intensities are fairly
consistent with the predictions of the SNR with $n_0$=0.01 atoms$\uden$,
$E_0=1.0\times10^{51}\ergs$, and $P_{nth}=7200\K\uden$ (blue dashed curve) at the age of 
$\sim$$10^5\yr$ (see the vertical dashed line in each panel). We also compare the SNR predictions with
those for the LTC of our BPL (horizontal dotted lines in each
panel with errors noted by the set of dotted lines at the top of right of the R12 and R45 panels;
the LTC's \ovi\ intensity and error bars are the same as those for the total BPL). The SNRs with
$n_0$=0.02 atoms$\uden$ (green curves) match those predicted by the LTC
at age $\sim1.8\times10^5\yr$ (see the vertical dotted line), but the SNR models
somewhat underestimate the LTC's R12 intensity (green solid and dotted lines) or overestimate its
R45 intensity (green dashed line).}
\label{fig:io6r1245}
\end{figure*}

\clearpage
\begin{deluxetable}{llcccccccc}
\tabletypesize{\scriptsize}
\tablewidth{0pt}
\tablecaption{$1T$ and $2T$ Halo Model}
\tablehead{
& \multicolumn{2}{c}{\underline{~~~~~~~~Local Bubble~~~~~~~}} &
\multicolumn{3}{c}{\underline{~~~~~~~Halo(cool)~~~~~~~~}} &
\multicolumn{3}{c}{\underline{~~~~~~~Halo(hot)~~~~~~~~~}} &\\
No.~~~Model &
$\log T$\tablenotemark{a,f} &
EM\tablenotemark{b,f} &
$\log T$\tablenotemark{a,f} &
EM\tablenotemark{b,f} &
$\tau$\tablenotemark{c,f} &
$\log T$\tablenotemark{a,f} &
EM\tablenotemark{b,f} &
$\tau$\tablenotemark{c,f} &
$\chi^2$/dof\\
}
\startdata
1~~~ $1T$~CIE(S+R)\tablenotemark{d} & $6.03^{+0.03}_{-0.04}$ & $ 8.9^{+0.4}_{-1.0}$ &
 $6.35^{+0.01}_{-0.01}$ & $11.8^{+1.5}_{-0.9}$ & \nodata & \nodata & \nodata & \nodata & 693.3/537\\

2~~~ $1T$~CIE(S)\tablenotemark{e} & $6.03$ & $ 8.9$ & $6.36^{+0.01}_{-0.01}$ & $11.5^{+0.5}_{-0.7}$ &
\nodata & \nodata & \nodata & \nodata & 591.2/535\\

3~~~ $1T$~NEI(S) & $6.03$ & $ 8.3$ & $6.37^{+0.01}_{-0.01}$ & $17.6^{+0.8}_{-1.0}$ &
$35.0^{+15.0}_{-33.2}$ & \nodata & \nodata & \nodata & 573.8/534\\

4~~~ $2T$~CIE(S+R)& $5.92^{+0.04}_{-0.05}$ & $7.2^{+0.6}_{-0.4}$ & $6.12^{+0.02}_{-0.01}$ &
$24.1^{+7.5}_{-3.4}$ & \nodata & $6.50^{+0.02}_{-0.02}$ & $5.6^{+2.6}_{-0.7}$ & \nodata & 567.3/535\\

5~~~ $2T$~CIE(S) & $5.92$ & $7.2$ & $6.16^{+0.01}_{-0.01}$ & $18.3^{+2.3}_{-4.3}$ &
\nodata & $6.51^{+0.03}_{-0.02}$ & $ 5.1^{+0.8}_{-0.6}$ & \nodata & 525.6/533\\

6~~~ $2T$~NEI+CIE(S) & $5.92$ & $7.2$ & $6.26^{+0.02}_{-0.03}$ & $14.0^{+2.3}_{-2.3}$ &
$0.306^{+0.116}_{-0.202}$ & $6.51$ & $5.1$ & \nodata & 524.3/534\\

7~~~ $2T$~CIE+NEI(S) & $5.92$ & $7.2$ & $6.16$ & $18.3$ &
\nodata & $6.52^{+0.03}_{-0.02}$ & $ 8.0^{+ 1.1}_{-1.1}$ & $1.36^{+ 48.6}_{-0.76}$ & 519.4/534\\

8~~~ $2T$~NEI+NEI(S)& $5.92$ & $7.2$ & $6.33^{+ 0.18}_{-0.10}$ & $9.9^{+ 3.1}_{-2.6}$ & 
$0.17^{+0.49}_{-0.14}$ & $6.52^{+0.04}_{-0.04}$ & $ 7.2^{+2.0}_{-1.9}$ & $ 31.7^{+18.3}_{-31.5}$ & 516.7/531\\
\enddata
\tablenotetext{a}{In unit of K.}
\tablenotetext{b}{In unit of $10^{-3}\uem$.}
\tablenotetext{c}{In unit of $10^{12}\utau$.}
\tablenotetext{d}{``(S+R)" means fit to \Suzaku+\ROSAT\ data simultaneously.}
\tablenotetext{e}{``(S)" means fit to \Suzaku\ spectra only.}
\tablenotetext{f}{The noted error bars reflect 90\% confidence intervals.}
\label{tab:2T model}
\end{deluxetable}

\begin{deluxetable}{lccc}
\tabletypesize{\small}
\tablewidth{0pt}
\tablecaption{Intrinsic Halo \ovii\ and \oviii\ Intensities}
\tablehead
{
Model & \Iovii\ & \Ioviii\ \\
& ($\lu$) & ($\lu$)
}
\startdata
2T\tablenotemark{a}    & $9.98^{+1.10}_{-1.99}$   & $2.66^{+0.37}_{-0.30}$\\
DM\tablenotemark{b}  & $10.6^{+0.6}_{-0.9}$ & $2.5^{+0.5}_{-0.3}$\\
\enddata
\tablenotetext{a}{Obtained form the CIE $2T$ model fitting to the \Suzaku\ spectra.}
\tablenotetext{b}{Direct measurement, described in Section 5.1.}

\label{tab:O emission}
\end{deluxetable}

\begin{deluxetable}{lcc}
\tabletypesize{\small}
\tablewidth{0pt}
\tablecaption{Intrinsic Halo \civ, \ovi, \ovii, and \oviii\ Intensities}
\tablehead
{
Ion & Average Energy & Intrinsic Intensity \tablenotemark{a}\\
& (eV) & ($\lu$)
}
\startdata
\civ   & $\sim$8.0   & $7780\pm2680$\\
\ovi   & $\sim$12    & $7750^{+950}_{-1090}$\\
\ovii  & $\sim$570 & $9.98^{+1.10}_{-1.99}$\\
\oviii & $\sim$650 & $2.66^{+0.37}_{-0.30}$\\
\enddata
\tablenotetext{a}{The \civ\ and \ovi\ error bars are 1$\sigma$. The \ovii\ and \oviii\ error bars reflect the
90\% confidence intervals.}
\label{tab:4 emission lines}
\end{deluxetable}

\begin{deluxetable}{lccc}
\tabletypesize{\small}
\tablewidth{0pt}
\tablecaption{Single Component Power-law Halo DEM model Patterned on the Model in \cite{shelton07}.}
\tablehead
{
$\log T_2$ & $\alpha$ & \Iovii & \Ioviii \\
(K) & & ($\lu$) & ($\lu$)
}
\startdata
6.06 & $1.09$ & $2.28$  & $0.00$\\
6.24 & $0.53$ & $10.4$  & $0.49$\\
6.36 & $0.37$ & $18.4$  & $2.97$\\
6.54 & $0.24$ & $22.9$  & $9.44$\\
\enddata
\label{tab:SPL model 1}
\end{deluxetable}

\begin{deluxetable}{lccccccc}
\tabletypesize{\small}
\tablewidth{0pt}
\tablecaption{Fitting a Single Component Power-law Halo DEM Model to the \Suzaku+\ROSAT\
Spectra.}
\tablehead{
\multicolumn{2}{c}{\underline{~~~~~~~~~Local Bubble~~~~~~~~}} &
\multicolumn{3}{c}{\underline{~~~~~~~~~~~~~~~~~~~Halo~~~~~~~~~~~~~~~~~~~}}&&&\\
$\log T$ & EM & $\log T_1$ & $\log T_2$ & $\alpha$ & $\Iovi$ & $\Iciv$ &
$\chi^2$/dof\\
(K) & ($10^{-3}\uem$) & (K) & (K) & & ($\lu$) & ($\lu$) &}
\startdata
$5.95^{+0.05}_{-0.06}$ & $ 6.4^{+0.5}_{-0.5}$ & $5.06$ & $6.61^{+0.03}_{-0.04}$ &
$-1.30^{+0.20}_{-0.16}$ & $46,800^{+2700}_{-5000}$ & $83,000^{+4800}_{-8800}$ & 538.9/536\\

$5.95^{+0.05}_{-0.05}$ & $ 6.4^{+0.5}_{-0.5}$ & $5.24$ & $6.61^{+0.04}_{-0.05}$ &
$-1.30^{+0.22}_{-0.15}$ & $46,500^{+17,700}_{-9930}$ & $7180^{+2740}_{-1530}$ & 538.9/536\\

$5.95^{+0.05}_{-0.05}$ & $ 6.4^{+0.6}_{-0.5}$ & $5.57$ & $6.61^{+0.02}_{-0.03}$ &
$-1.30^{+0.21}_{-0.19}$ & $7850^{+1400}_{-1160}$ & $181^{+32}_{-27}$ & 539.2/536\\

$5.94^{+0.05}_{-0.06}$ & $6.6^{+ 0.5}_{-0.5}$ & $5.76$ & $6.61^{+0.05}_{-0.03}$ &
$-1.38^{+0.27}_{-0.29}$ & $861^{+57}_{-525}$ & $38.0^{+2.5}_{-23.2}$ & 541.9/536\\

\enddata
\label{tab:SPL model 2}
\tablecomments{For each fit, $\log T_1$ is fixed at the specified value, and the other halo and LB
parameters are free to vary. The \ovi\ and \civ\ intensities are then derived from the best-fit model
parameters. The $\chi^2/dof$ pertains to the fit to the \Suzaku+\ROSAT\ data.}
\end{deluxetable}

\begin{deluxetable}{lcccccccc}
\tabletypesize{\scriptsize}
\tablewidth{0pt}
\tablecaption{Fitting Broken Power-law Halo DEM Models to the \Suzaku+\ROSAT\ Spectra.}
\tablehead{
&
\multicolumn{2}{c}{\underline{~~~~~~~~~Local Bubble~~~~~~~~~}} &
\multicolumn{2}{c}{\underline{~~~Halo(LTC)~~~}} &
\multicolumn{2}{c}{\underline{~~~~~~~Halo(HTC)~~~~~~~}} &\\
$\log T_2$& $\log T$ & EM &
$\log T_1$ & $\alpha_1$ & $\alpha_2$ &
$\log T_3$ & \Iciv & $\chi^2$/dof\\
(K) & (K) & ($10^{-3}\uem$) & (K) &  & (K) & & ($\lu$) &
}
\startdata
5.97 & $5.93^{+0.04}_{-0.05}$ &  $ 6.8^{+ 0.5}_{-0.5}$ & 4.8 & 0.00 & $-2.10^{+0.11}_{-0.19}$ &
$6.95^{+0.02}_{-0.02}$ & $9500$ & 544.2/536\\

6.02 & $5.92^{+0.02}_{-0.05}$ & $ 6.8^{+ 0.5}_{-0.4}$ & 4.8 & 0.30 & $-2.21^{+0.19}_{-0.12}$ &
$7.02^{+0.03}_{-0.05}$ & 7440 & 541.6/536\\

6.06 & $5.93^{+0.05}_{-0.05}$ & $6.8^{+ 0.5}_{-0.6}$ & 4.8 & 0.38 & $-2.06^{+0.39}_{-0.11}$ &
$6.96^{+0.04}_{-0.18}$ & $6860$ & 541.8/536\\

6.14 & $5.88^{+0.04}_{-0.04}$ & $ 6.5^{+ 0.7}_{-0.6}$ & 4.8 & 0.54 & $-1.58^{+0.41}_{-0.17}$ &
$6.94^{+0.01}_{-0.03}$ & $ 5800$ & 590.2/536\\

\enddata
\tablecomments{For each fit, $\log T_1$ and $\log T_2$ are fixed at the specified value, and the other
halo and LB parameters are free to vary. The \civ\ intensities are derived from the best-fit model parameters.
The values of $\chi^2$ in the final column are obtained by fitting the model
to the \Suzaku+\ROSAT\ data.}
\label{tab:BPL model}
\end{deluxetable}

\begin{deluxetable}{lcccccccccc}
\tabletypesize{\scriptsize}
\tablewidth{0pt}
\tablecaption{Broken Power-law Halo Model: Testing Different Abundance Tables.}
\tablehead{
& \multicolumn{2}{c}{\underline{~~~~~~~~Local Bubble~~~~~~~}} &
\multicolumn{3}{c}{\underline{~~Halo(LTC)~~}} &
\multicolumn{2}{c}{\underline{~~~~~~~Halo(HTC)~~~~~~~~~}} & & &\\
Table & $\log T$ & EM & $\log T_1$ & $\log T_2$ &
$\alpha_1$ & $\alpha_2$ & $\log T_3$ & \Iovii & \Ioviii & $\chi^2$/dof\\
& (K) & ($10^{-3}\uem$) & (K) & (K) &  &  & (K) & ($\lu$) & ($\lu$) &
}
\startdata

Grsa\tablenotemark{a} & $5.95^{+0.05}_{-0.05}$ & $ 4.2^{+ 0.3}_{-0.3}$ &  4.8 & 6.06 & 0.11 &
$-2.16^{+0.12}_{-0.15}$ & $6.96^{+0.00}_{-0.04}$ & $11.1^{+2.4}_{-1.5}$ & $2.84^{+0.63}_{-0.42}$ &
532.8/536\\

Lodd\tablenotemark{b} & $5.95^{+0.05}_{-0.06}$ & $ 4.7^{+ 0.4}_{-0.3}$ & 4.8 & 6.06 & -0.25 &
$-2.40^{+0.12}_{-0.18}$ & $6.96^{+0.00}_{-0.02}$ & $9.67^{+2.60}_{-1.63}$ & $2.49^{+0.58}_{-0.42}$ &
541.1/536\\

Wilm\tablenotemark{c} & $5.93^{+0.05}_{-0.05}$ & $ 6.8^{+ 0.5}_{-0.6}$ & 4.8 & 6.06 & 0.38 &
$-2.06^{+0.39}_{-0.11}$ & $6.96^{+0.04}_{-0.18}$ & $9.69^{+2.54}_{-2.98}$ & $2.47^{+0.82}_{-0.94}$ &
541.8/536\\

Aneb\tablenotemark{d} & $5.92^{+0.04}_{-0.05}$ & $ 4.1^{+ 0.3}_{-0.3}$ & 4.8 & 6.06 & 0.35 &
$-2.02^{+0.16}_{-0.19}$ & $7.06^{+0.00}_{-0.02}$ & $10.8^{+2.5}_{-1.6}$ & $2.86^{+0.71}_{-0.54}$ &
543.0/536\\

Feld\tablenotemark{e} & $5.94^{+0.04}_{-0.05}$ & $ 4.5^{+ 0.3}_{-0.3}$ & 4.8 & 6.06 & 0.72 &
$-1.70^{+0.20}_{-0.17}$ & $7.16^{+0.00}_{-0.03}$ & $10.6^{+2.6}_{-1.5}$ & $ 3.00^{+0.63}_{-0.44}$ &
554.1/536\\
 
Angr\tablenotemark{f} & $5.97^{+0.04}_{-0.04}$ & $ 4.3^{+ 0.3}_{-0.3}$ & 4.8 & 6.06 & 0.60 &
$-2.05^{+0.12}_{-0.15}$ & $7.06^{+0.00}_{-0.04}$ & $11.1^{+2.7}_{-1.7}$ & $2.81^{+0.69}_{-0.52}$ &
 563.3/536\\

\enddata
\label{tab:abun BPL model}
\tablecomments{For each fit, $\log T_1$ and $\log T_2$ are fixed at 4.8 and 6.06, respectively, and the
other halo and LB parameters are free to vary. The \ovii\ and \oviii\ intensities are then derived from the
best-fit model parameters. The values of $\chi^2$ in the final column are obtained by fitting the model
to the \Suzaku+\ROSAT\ data.}

\tablenotetext{a}{\cite{grevesse98}}
\tablenotetext{b}{\cite{lodders03}}
\tablenotetext{c}{\cite{wilms00}, except that the XSPEC version sets several elemental abundance to 0.}
\tablenotetext{d}{\cite{anders82}}
\tablenotetext{e}{\cite{feldman92}}
\tablenotetext{f}{\cite{anders89}}
\end{deluxetable}

\begin{deluxetable}{lccc}
\tabletypesize{\small}
\tablewidth{0pt}
\tablecaption{Broken Power-law Halo Model: Predicted Soft X-ray Count Rates}
\tablehead
{
Component & R12($1/4\kev$)\tablenotemark{a} & R45($3/4\kev$)\tablenotemark{a} &
R67($1.5\kev$)\tablenotemark{a}
}
\startdata
LB\tablenotemark{b} & $417^{+126}_{-75}$ & $0.427^{+0.129}_{-0.077}$ & $0.00353^{+0.00107}_{-0.00064}$\\
Low T Component\tablenotemark{c}  & $339^{+103}_{-61}$ & $4.77^{+1.45}_{-0.86}$ & $0.186^{+0.056}_{-0.033}$\\
High T Component\tablenotemark{c}  & $284^{+86}_{-51}$ & $73.2^{+22.2}_{-13.2}$ & $20.4^{+6.2}_{-3.7}$\\
EPL\tablenotemark{c}& $57.7^{+17.5}_{-10.4}$ & $54.8^{+16.6}_{-9.9}$ & $104^{+32}_{-19}$\\
Total & $1098^{+333}_{-198}$ & $133^{+40}_{-24}$ & $124^{+38}_{-22}$\\
\enddata
\tablenotetext{a}{All values are in unit of $10^{-6}$~ROSAT~$\textnormal{counts}~\textnormal{s}^{-1}~\textnormal{arcmin}^{-2}$.}
\tablenotetext{b}{The calculated SXR count rates for this component are unabsorbed.}
\tablenotetext{c}{The calculated SXR count rates for this component have been subjected to absorption
due to $\nh=1.9\times10^{20}\pcms$.}
\label{tab:BPL model prediction}
\end{deluxetable}


\end{document}